\documentclass[pdflatex,sn-basic]{sn-jnl}
\usepackage{amsmath}
\usepackage{graphicx}%
\usepackage{multirow}%
\usepackage{amsmath,amssymb,amsfonts}%
\usepackage{amsthm}%
\usepackage{mathrsfs}%
\usepackage[title]{appendix}%
\usepackage{xcolor}%
\usepackage{textcomp}%
\usepackage{manyfoot}%
\usepackage{booktabs}%
\usepackage{algorithm}%
\usepackage{algorithmicx}%
\usepackage{algpseudocode}%
\usepackage{listings}%
\usepackage[normalem]{ulem}%

\begin{document}

\title[Article Title]{Exo-Geoscience Perspectives Beyond Habitabilty}


\author*[1]{\fnm{Tilman} \sur{Spohn}}\email{tilman.spohn@dlr.de}

\author[2]{\fnm{Aki} \sur{Roberge}}\email{}

\author[3,4]{\fnm{M. J.} \sur{Way}}
\author[5]{\fnm{Jo\~ao C.} \sur{Duarte}}
\author[6]{\fnm{Francesca} \sur{Miozzi}}
\author[7]{\fnm{Philipp} \sur{Baumeister}}

\author[8]{\fnm{Paul} \sur{Byrne}}
\author[9]{\fnm{Charles} \sur{Lineweaver}}
\affil*[1]{\orgdiv{Institute of Space Research}, \orgname{German Aerospace Center (DLR)}, 
\city{Berlin},  
\country{Germany}}

\affil[2]{\orgdiv{Astrophysics Division}, \orgname{NASA Goddard Space Flight Center}, 
\city{Greenbelt}, 
\country{USA}}

\affil[3]{\orgname{NASA Goddard Institute for Space Studies}, 
\city{New York}, 
\country{USA}}

\affil[4]{\orgdiv{Theoretical Astrophysics}, \orgname{Department of Physics and Astronomy}, \city{Uppsala}, 
\country{Sweden}}

\affil[5]{\orgdiv{Instituto Dom Luiz, Faculty of Sciences}, 
\orgname{University of Lisbon}, \city{Lisbon}, 
\country{Portugal}}

\affil[6]{\orgdiv{Earth and Planets Laboratory}, \orgname{Carnegie Institution for Science}, 
\city{Washington D.C.}, \country{United States}}

\affil[7]{\orgdiv{Department of Earth Sciences}, \orgname{FU Berlin}, 
\city{Berlin}, \country{Germany}}

\affil[8]{\orgdiv{Department of Earth, Environmental, and Planetary Sciences}, \orgname{Washington University}, 
\city{St Louis}, \country{United States}}

\affil[9]{\orgdiv{Planetary Science Institute}, \orgname{The Australian National University}, 
\city{Canberra}, \country{Australia}}


\abstract{This article reviews the emerging field of exo-geoscience, focusing on the geological and geophysical processes thought to influence the evolution and (eu)habitability of rocky exoplanets. We examine the possible roles of planetary interiors, tectonic regimes, continental coverage, volatile cycling, magnetic fields, and atmospheric composition and evolution in shaping long-term climate stability and biospheric potential. Comparisons with Earth and other planets in the Solar System highlight the  diversity of planetary conditions and the rarity of conditions relevant to life. We also discuss contingency and convergence in planetary and biological evolution as they relate to the spread of life in the universe. The observational limits of current and planned missions are assessed, emphasizing the need for models that connect internal dynamics to detectable atmospheric and surface signatures as well as the need for laboratory measurements of planetary properties under a wide range of conditions. The large number of exoplanets promises opportunities for empirical and statistical studies of processes that may have occurred earlier in Earth's history, as well as of the other pathways rocky planets and biospheres may take. Thus, exo-geoscience provides a framework for interpreting exoplanet diversity and refining strategies for detecting life beyond the Solar System.}


\maketitle

\section{Introduction}\label{sec:intro}

``Where is everybody?", Enrico Fermi is reported to have asked Emil Konopinski, Edward Teller, and Herbert York on their way to lunch at the Los Alamos National Laboratory some time in the summer of 1950 \citep{jones_fermi_1985}. The question verbalizes what has become known as Fermi's paradox \citep[e.g.,][]{webb2015if}: In an universe of presumably trillions of Earth-like planets, why have we not seen signs of (intelligent) life beyond the Earth? Seventy-five 
years later, the question still puzzles scientists and the interested public, although the paradox has been discussed numerous times \citep[e.g.,][]{webb2015if}. 
Instead, it has recently been argued on the basis of Bayesian statistics and observational data that our planet may be rare and that the vast majority of formally habitable planets in the galaxy may lack important elements to make them truly habitable and inhabited \citep[e.g.,][]{simpson_bayesian_2017, scherf_eta-earth_2024,apai2025terminology}.  

Although exoplanet research is justified in its own right and as part of the broader field of comparative planetology, our focus here is on exoplanets as potential hosts of extraterrestrial life. Over 6000 exoplanets have been detected, with another almost 8000 candidates awaiting confirmation 
(\url{https://exoplanetarchive.ipac.caltech.edu}, retrieved Dec 17, 2025). 
This includes an increasing number of planets that are approximately the size and mass of Earth. A significant number of these planets orbit within their host stars' habitable zone\footnote{The circumstellar habitable zone is the annular region around a star where the physical conditions allow the existence of liquid water on the surface of a rocky planet \citep[e.g.,][section 8.1]{lingam_stars_2024}. It was first introduced by \cite{hart_habitable_1979} and quantified by \cite{Kasting:1993}.} 
and may have the potential for liquid surface water. While the concept of the habitable zone is useful for a first--order assessment of the chance of detecting extraterrestrial life, further refinement is likely needed before investing in costly and sophisticated follow-up observations \citep[e.g.,][]{glaser_detectability_2020}. More specific planetary physical properties to consider include the balance between land and (deep) ocean surfaces, the availability of water versus nutrients from riverine sources such as phosphorus and nitrogen, signs of thermodynamic disequilibrium \citep[e.g.][]{schwieterman2018}, and the potential for building-up free oxygen to support more complex life forms. In addition, tectonic modes and magnetic fields may matter. The article by \cite{glaser_euhabitable_2026} in this topical collection introduces a new term to capture these additional factors: ``euhabitable'', meaning \textit{truly} habitable. 
While these planetary properties are difficult to observe remotely with present means (see Section \ref{sec:observability} below), they may be important for understanding the conditions under which life can form and evolve.

\begin{figure}
    \centering
    \includegraphics[width=0.75\linewidth]{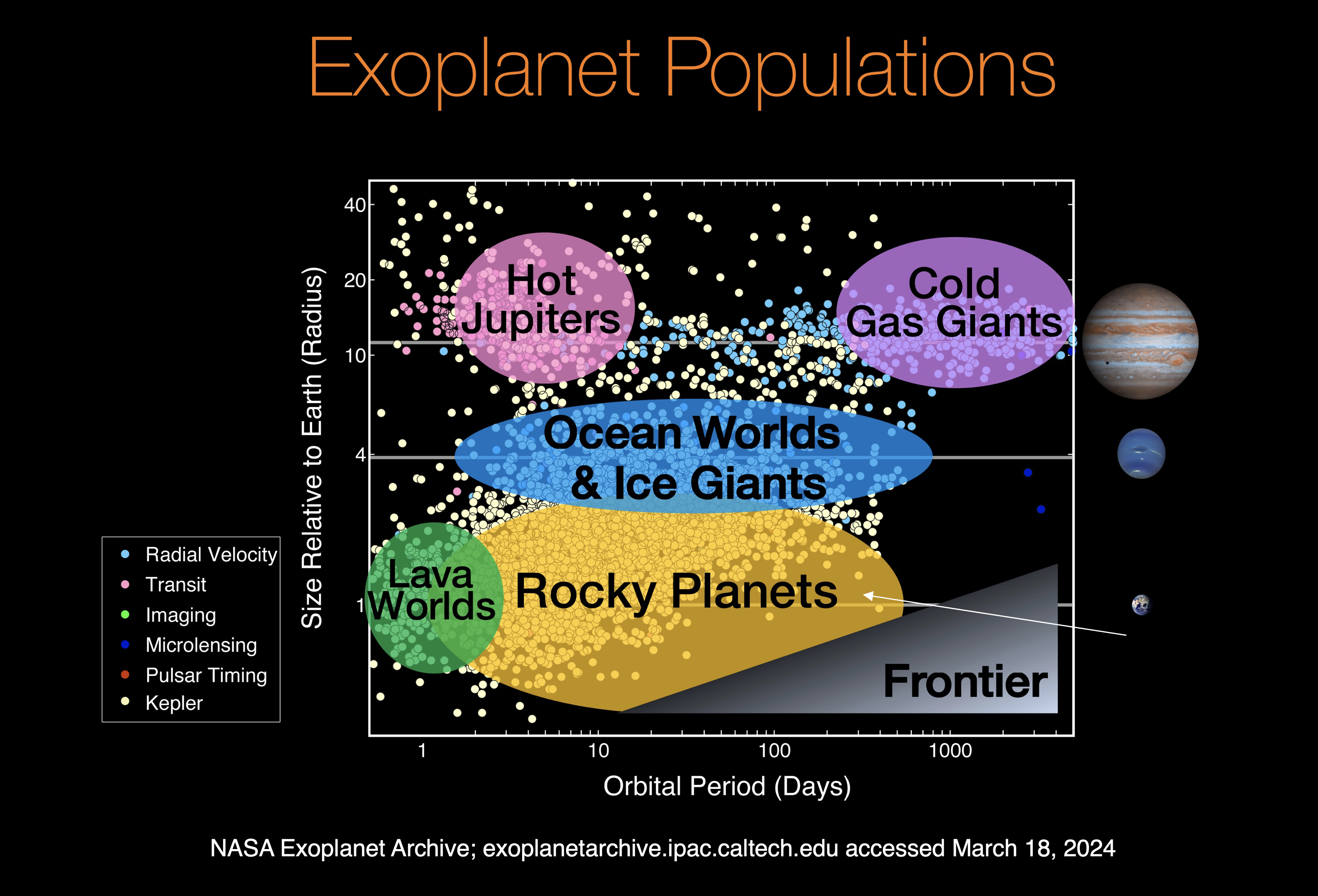}
    \caption{The NASA Exoplanet Archive compiled exoplanet populations. The legend at the left--hand side indicates the detection methods. To date, 6065 exoplanet detections have been confirmed, and   7687 candidates are awaiting confirmation. Of the confirmed planets, 1446 have radii  of 2 Earth radii or less, masses between 0.5 and 10 Earth masses, and estimated densities between 3000 and 6000 kg/m$^3$.
    ``Rocky planets'', the subject matter of this topical collection, include a wide variety of planets -- from sub- to superterran sizes, with or without atmospheres, and with presumably a wide variety of surfaces and interior structures (\url{https://exoplanetarchive.ipac.caltech.edu}, retrieved Dec 17, 2025).
}
    \label{fig:exo-populations}
\end{figure}

For geoscientists, rocky exoplanets allow for an assessment of how unusual or common Earth is, and motivating scientists to reconsider Earth's properties in new ways. To quote \cite{shorttle_geosciences_2021}, “\ldots many of Earth’s one-offs -- plate tectonics, surface liquid water, a large moon, and life: long considered as `Which came first?' conundrums for geoscientists -- may find resolution in the study of exoplanets that possess only a subset of these phenomena.”  Eventually sufficient numbers of exoplanets may be observed at different stages in their evolution to allow statistically robust conclusions to be drawn. We may also have the chance to observe planets that took different evolutionary paths, 
whether geological or biological. The search for life beyond the Solar System may even further our understanding of the future of life on Earth, albeit indirectly, as discussed in e.g., \citet[][Part IV]{lingam_stars_2024}. For example, we may learn whether complex life persists once it arises or whether such life represents a fleeting moment in a planet's evolution on geological timescales.     

This article is organized into three main sections. The next section discusses what can be inferred from studies of Earth and terrestrial life for exoplanet and life detection studies. Next, we report on upcoming missions to observe rocky planets, with an emphasis on life detection. Lastly, we discuss the interest that geoscientists may have in the diversity of planets. We close by sharing some concluding thoughts on the emerging field of Exo-Geoscience.

\section{Geoscience Lessons for Exoplanet Science}\label{sec2}

The Copernican Principle\footnote{The Copernican Principle states that Earth does not occupy a special or privileged position in the universe.} \citep[e.g.,][]{bondi_cosmology_1952} and the Principle of Mediocrity\footnote{The Principle of Mediocrity states that Earth is not special or unique within the cosmos.} - both widely accepted - form the basis of the argument that the universe should be teeming with life (for a recent review and critical discussion of the mediocrity principle as applied to astrobiology, see \citealt{balbi_beyond_2023} and \citealt{lingam_stars_2024}). However, from the perspective of the Solar System, Earth appears unique as the only planet known to support life. Its combination of geological and astronomical properties may even be rare in the present galaxy, perhaps even in the Universe \citep[e.g.,][]{taylor_destiny_1998, ward_rare_2000, waltham2014lucky, simpson_bayesian_2017, scherf_eta-earth_2024}. Some characteristics that may be particularly important but rare are: 
\begin{itemize}
    \item the right size (mass) rocky planet in the star's habitable zone 
    \item plate tectonics and a magnetic field, 
    \item land masses and oceans on the surface,
    \item a N$-$O atmosphere with some CO$_2$, or simply N+CO$_2$ 
    \item a large moon stabilizing the rotation axis against obliquity excursions and thus stabilizing climate zones  \citep{laskar1993stabilization}, and
    \item a giant planet like Jupiter guarding against a high impact flux \citep[e.g.][]{wetherill1994possible}.   
\end{itemize}
Other properties are more related to the Earth's place in the galaxy, such as orbiting a G-type star in the galactic habitable zone\footnote{The galactic habitable zone GHZ \citep[e.g.,][]{gowanlock2011model,lineweaver2025galactic} is defined as the region of space and time within a galaxy that is compatible with life-as-we-know-it. These regions contain main-sequence stars that are relatively far from life-extinguishing supernovae. Main-sequence stars in the GHZ have sufficient metallicity to allow the formation of rocky planets, yet not enough to produce giant planets that can migrate through the circumstellar habitable zone. The main-sequence lifetimes of these stars should allow for the formation and evolution of life.}, and perhaps early ejection of the Sun from its original stellar nursery \citep[e.g.][]{adams2010birth}.

The Rare Earth hypothesis has been criticized for being anthropocentric and overemphasizing the needs of \textit{life-as-we-know-it}. For example, \citet{kasting_Rare_Earth_2001} and specifically \citet[][p.384]{langmuir_habitable_2012} argue that life  is an ``efficient and natural planetary process'', drawing energy from 
its environment while increasing the system's entropy and maximizing entropy production \citep[see also][]{kleidon2004non}. They argue that this process should ``occur widely throughout the universe''. 
Others, such as \cite{southam_geology_2015}, argue that water and rock should provide all the ingredients needed for life. Indeed, we may be too focused on life as we know it; to quote (not verbatim) the Greek philosopher Xenophanes (570 BC), ``if horses had gods, they would represent them as horses''  (see also \citealt{feinberg1980life} and \citealt{ward2007life}).  But how can we design a search for life that is completely different from what we know? Without a deep understanding of life in general — how it originates and evolves — and how alien life forms and their basis might differ from those on Earth, it is only natural to search for habitats similar to those on our own planet and the other planets in the Solar System.

Commonly used terms such as habitable -- or more recently introduced terms such as urable \citep[conditions that allow life to begin;][]{deamer_urability_2022}, euhabitable (truly habitable; \cite{glaser_euhabitable_2026}, this topical collection), and fecundity \citep[suitable to support life with technological intelligence\footnote{We distinguish between intelligent life and life with technological intelligence, since animals can possess intelligence.};][]{simpson_bayesian_2017} -- thus mostly reflect life as we know it (see also \cite{apai2025terminology} for a discussion of relevant terminology). 
One open question is how likely it is that a urable and (eu)habitable planet, or even a planet with fecundity, would actually be inhabited. As  \citet{deamer_urability_2022} argue, a urable planet would almost certainly exhibit abiogenesis. Due to the proximity of the conditions for urability and euhabitability, it is reasonable to suppose that such a planet would have a high probability of being (or having been) inhabited. However, note that an (eu)habitable planet can evolve away from urability because the latter requires an anoxic environment\footnote{Molecular oxygen is highly reactive, so the reduced organic compounds necessary for the origin of life would be degraded by oxidation \citep[e.g.,][]{deamer_urability_2022, rimmer2021life}.}.  Thus, Earth's surface today -- except for certain anoxic niches -- would not qualify as urable (see \citealt{deamer_urability_2022} and \citealt{lineweaver_habitability_2012} for discussions). This observation let \cite{lineweaver2025other} to propose that life must escape a ``bottleneck" before it can coevolve with the planet.

Many features of the present-day Earth (ignoring anthropogenic features) can be linked to the biosphere in one way or another. These features include the composition of the atmosphere \citep[e.g.,][]{grenfell_review_2017}, oceans \citep[e.g.,][]{langmuir_habitable_2012}, and soil \citep[e.g.,][]{retallack_coevolution_2015}, as well as processes like plate tectonics, continental growth \citep[e.g.,][]{rosing_rise_2006, retallack_coevolution_2015, spencer2022biogeodynamics, honing_land_2023, stern_2024}, and mountain building \citep[e.g.,][]{parnell_increased_2021}, which have all been linked to bioactivity. Microbial activity, moreover,  drives Earth's biogeochemical cycles and promotes the oxygenation of the planet \citep{falkowski2008microbial}.  Consequently,  
\cite{zuluaga2014habitable} have proposed the terms ``abiotic" and ``inhabited habitable zones" to acknowledge that physical parameters used to delineate the boundaries of a circumstellar habitable zone may be modified by life. From a chemical perspective, life -- as \cite{langmuir_habitable_2012} noted -- is a complement of the solid Earth.

\subsection{The Earth and the Solar System planets as reference cases: Properties and processes}\label{subsec2.1}

Earth is the $3^{rd}$ planet from the Sun in the Solar System.  
The Earth in comparison with Solar System planets and moons and known exoplanets is the subject of the article by \citet{byrne2026} in this topical collection. The mass of Earth is sufficient to gravitationally bind a substantial N--O--dominated atmosphere (see \citet{Steinmeyer2026} and \citet{kubyshkina2026}, this topical collection). Venus, of similar mass, binds a much heavier, CO$_2$-dominated atmosphere with only minute amounts of molecular oxygen \citep[e.g.,][]{baines2015atmospheres}. It is widely believed that the difference in atmospheric composition between the two neighboring planets is caused by differences in their early evolution \citep[e.g.][]{Hamano2013,lebrun2013} or was driven by a yet to--be--determined catastrophic climate change on Venus sometime in its history \citep{way2020,way2022large}. 
Mars, about half the size of Earth and 1/10th the mass, could not sustain a habitable atmosphere long enough for a sustained surface life cycle \citep{Wordsworth2016}, so any remaining biosphere would be subterranean.

\subsubsection{Interior properties and processes}

\begin{figure}
     \centering
     \includegraphics[width=0.75\linewidth]{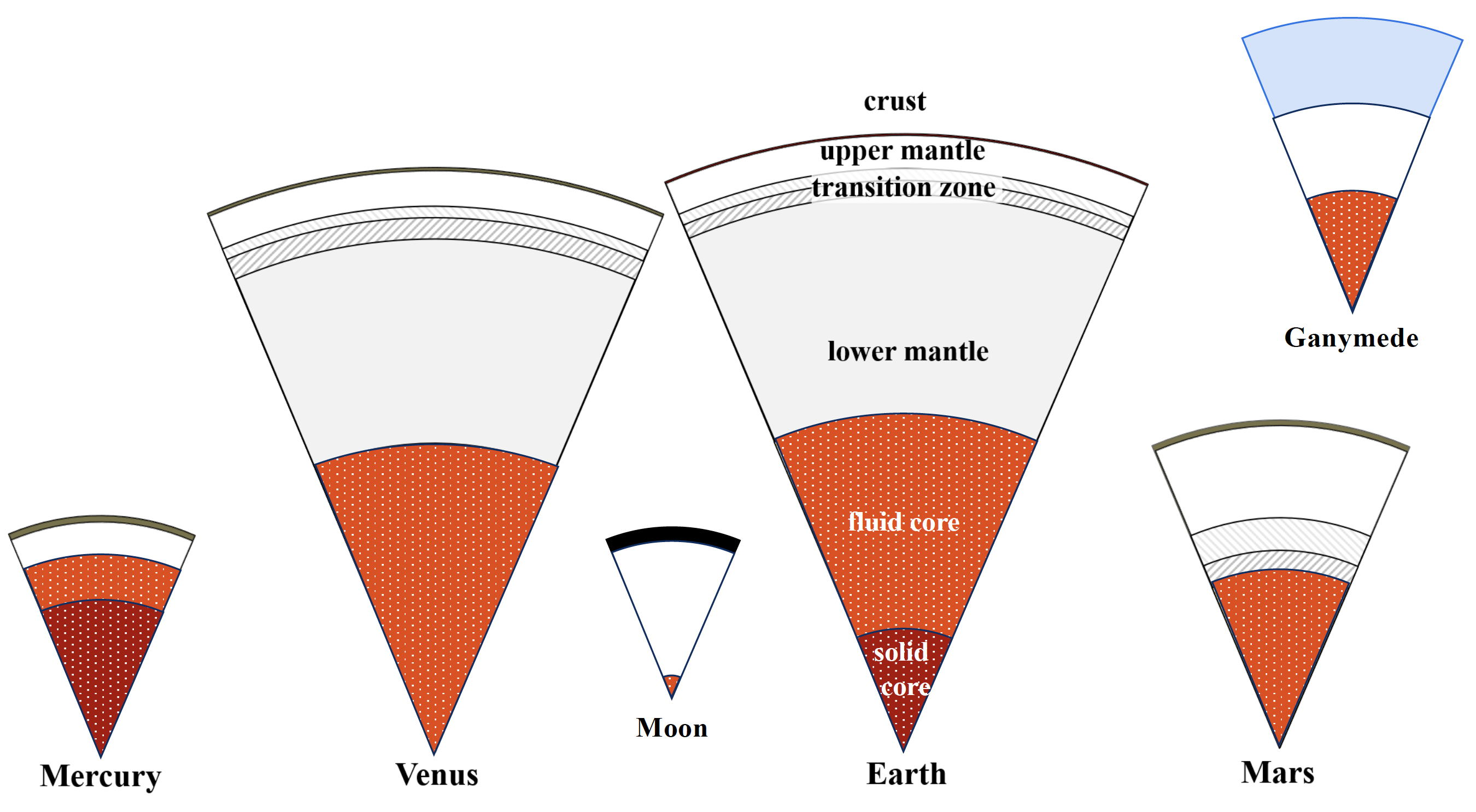}
     \caption{Interior structures of representative Solar System objects. The Earth, Venus, and Mars have iron-rich cores (red sections) with radii about half that of the planet. This is typical of other differentiated Solar System objects as well. For icy Solar System satellites, the ratio of 0.5 applies to the core radius relative to the satellite radius minus the ice shell thickness (blue).  Mercury and the Moon have anomalous core-to-planetary radius ratios of approximately 0.8 and 0.2, respectively. Planetary cores are generally molten initially but may freeze as the planets cool. Earth and Mercury have solid inner cores (dark red). Venus, Mars, and Ganymede may have solid inner cores as well, although the absence of magnetic fields at Venus and Mars suggests that their cores may still be fully molten \citep{Stevenson:1983}. The rocky mantles of Earth and Venus are compositionally layered with phase transitions (olivine-spinell and $\beta$ to $\gamma$ spinel) dividing the layers. These high-pressure phase transformations likely also occur in the lowermost Martian mantle. (Modified after \citealt{breuer2023terrestrial})}
     \label{fig:structure}
 \end{figure}

\begin{figure}
    \centering
    \includegraphics[width=1\linewidth]{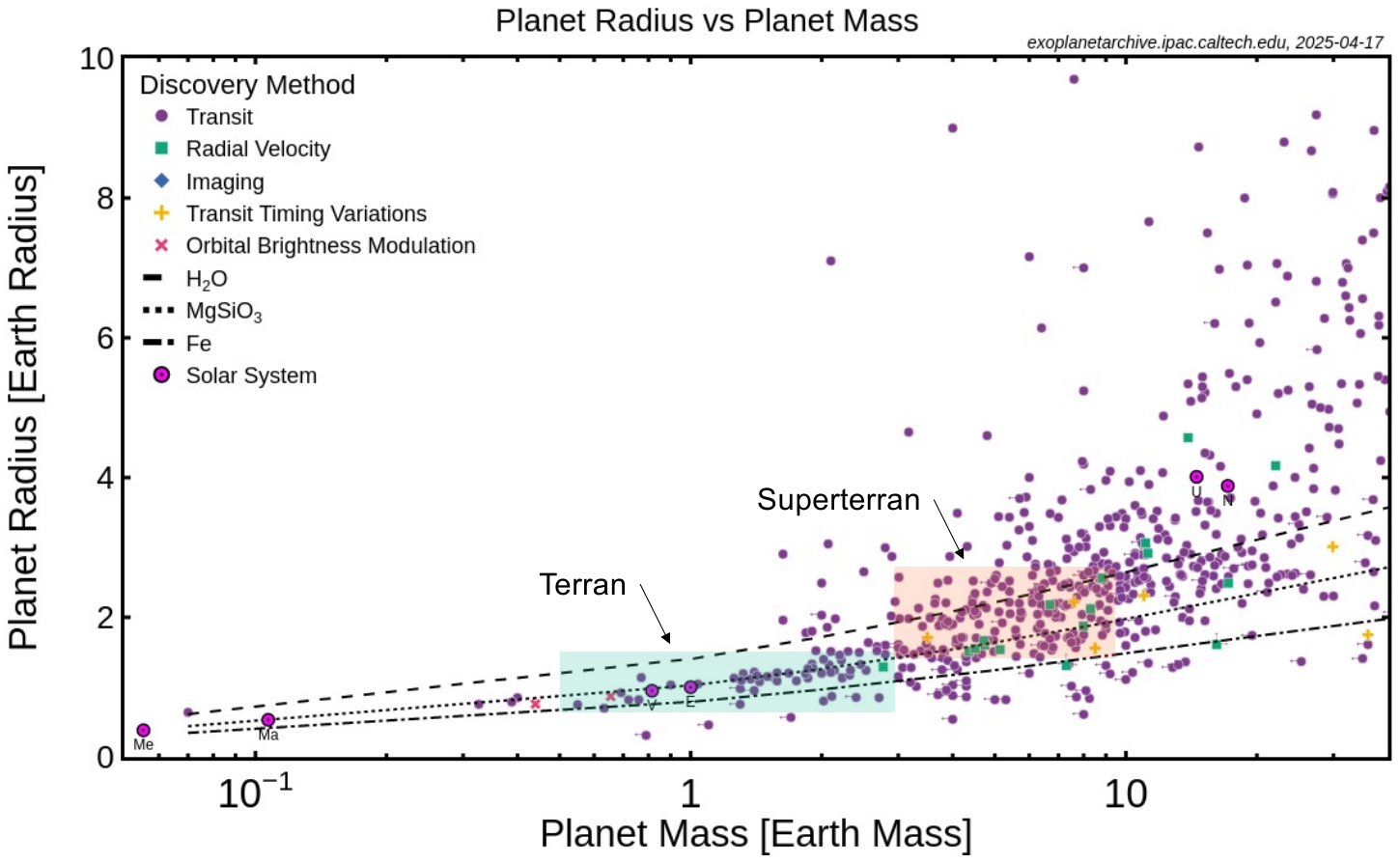}
    \caption{Radius versus mass for confirmed exoplanets as collected by Nasa's Exoplanet Archive (https://exoplanetarchive.ipac.caltech.edu, 2025-04-17). The different observation methods are indicated. Also plotted are the masses and radii of Solar System planets, the radii of model planets composed of H$_2$O, MgSiO$_3$ and Fe, and the fields of terran and superterran planets. Note that generally errors are large, in particular in mass. 
    Note that Venus, Earth, and Mars plot close to the MgSiO$_3$ curve.  See also Figures 1 and 2 of \cite{lineweaver2025other}
}
    \label{fig:mass-radius}
\end{figure}

\paragraph{Interior structure} \label{p:Interior structure}
Earth's interior is divided into a rocky mantle (and crust) and an iron-rich core that freezes over time by growing a solid inner core \citep[e.g.,][]{hirose2013composition}. The buoyancy released by freezing is widely agreed to drive a dynamo that generates Earth's magnetic field (compare section \ref{p:magnetism}). The radius of the core is about half the radius of the planet, a characteristic shared by most rocky planets and satellites in the Solar System with the notable exceptions of Mercury and the Moon (compare Fig. \ref{fig:structure}). Even for the ice-rich satellites of the outer Solar System, the ratio of the iron--core radius to the satellite radius minus the thickness of the ice shell is about 0.5. The mass--radius statistics of exoplanets (Fig. \ref{fig:mass-radius}) suggest that this ratio is more variable for rocky exoplanets. Earth's solid inner core has grown over time to a radius of a little more than a third of the core radius. Geodetic data from the Messenger mission suggest that Mercury likely has a substantial solid inner core of 0.3 -- 0.7 core radii \citep{genova2019geodetic}. Solid Earth's outermost layer,  the crust, encompasses both the oceanic crust that underlies the oceans and the predominantly sub--aerial continental crust that constitutes the continents and continental shelves. 

The interior structure of Earth, and to a lesser extent Mars and the Moon, is well understood thanks to seismological exploration \citep[e.g.,][]{dziewonski_deep_2015, lognonne2015planetary,lognonne2019seis}. However, the interior structure of other rocky worlds in the Solar System is much less constrained by mass, radius, gravity field, and tidal deformation \citep[e.g.,][]{sohl_interior_2015}. This is even more true for rocky exoplanets \citep{baumeister2025fundamentals}. To infer the interior properties, in addition to mass and radius, the composition of the exoplanet's central star is used, which is justified by the similarity of the compositions of the Solar System bodies with that of the Sun \citep[e.g.,][]{palme_solar_2014}. The interior structure may indirectly influence (eu)habitability through its effect on tectonic styles, such as plate tectonics, and degassing and planetary magnetism, as discussed below.

\begin{figure}
    \centering
    \includegraphics[width=1.0\linewidth]{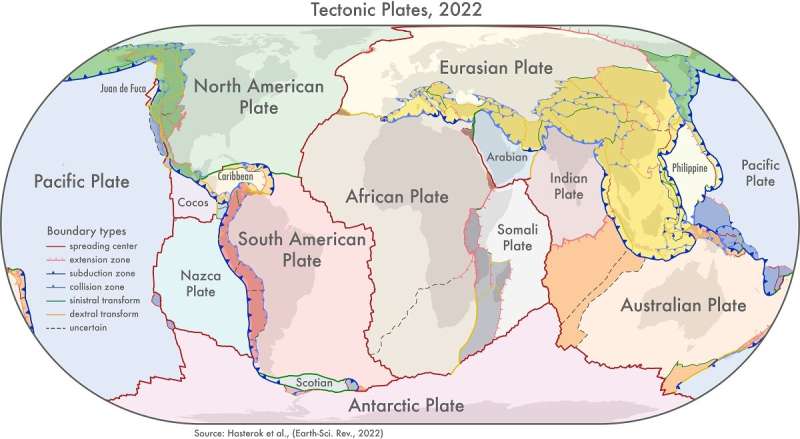}
    \caption{The tectonic plates of the Earth and their boundaries. There are seven major (African, Antarctic, Australian, Eurasian, North and South American, and Pacific) and nine minor plates.  Source: \url{http://phys.org/news/2022-06-global-geological-provinces-tectonic-plates.html} by D. Hasterok, University of Adelaide, based on \cite{hasterok2022new}. Retrieved July 21, 2025}
    \label{fig:tectonic-plates}
\end{figure}

\begin{figure}
    \centering
    \includegraphics[width=1\linewidth]{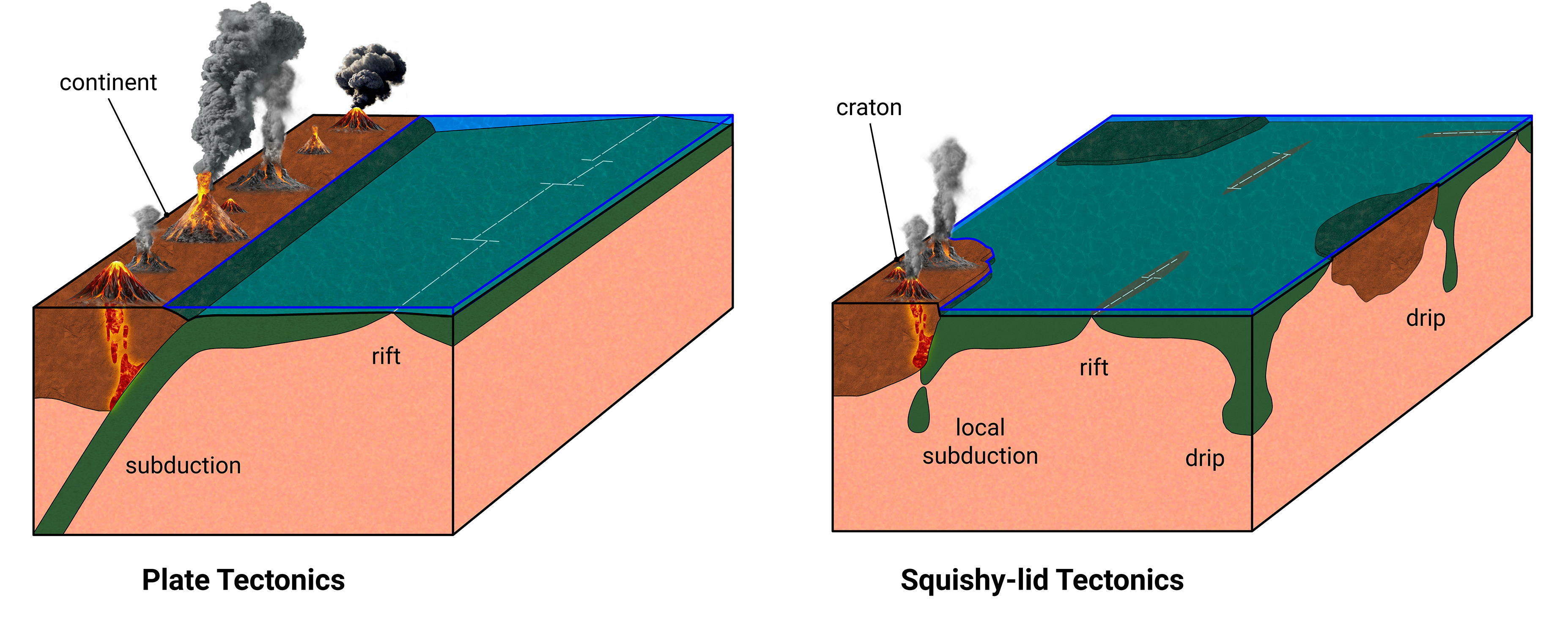}
    \caption{Schematic diagram illustrating modern plate tectonics and continental growth by subduction-related volcanism, as well as earlier squishy lid tectonics. Squishy lid tectonics may have predated plate tectonics during the early evolution of the Earth in the Archean when the lithosphere was warmer and more deformable. This, in turn, may have been predated by heat-pipe tectonics in the Hadean and possibly the early Archean. In this model, a stagnant lid is pierced by volcanic upwelling, and the mass balance is maintained by lithosphere delamination. Transitions between heat pipe, squishy lid, and plate tectonics were likely gradual.  (modified after \citealt{cawood_secular_2022})}
    \label{fig:plate vs squishy}
\end{figure}

\begin{figure}
    \centering
    \includegraphics[width=0.75
    \linewidth]{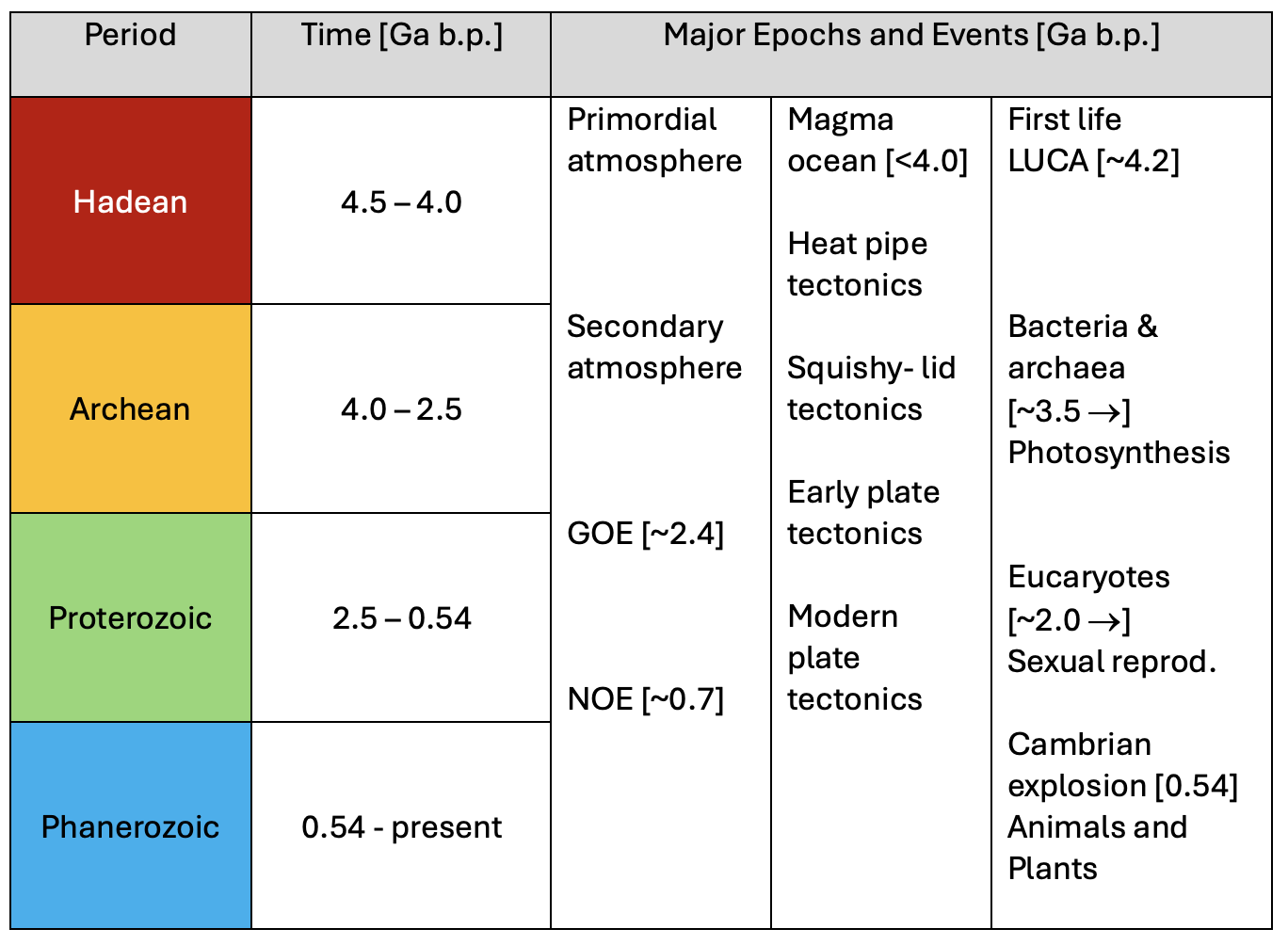}
    \caption{Simplified geological time table and major epochs and events used and discussed in the present paper. GOE denotes the great oxdation event and NEO the Neoproterozoic oxidation event}
    \label{fig:Geological time table}
\end{figure}

\paragraph{Plate tectonics and continents} \label{p:Plate tectonics and continents}
Heated by radiogenic elements and losing heat from accretion (the present--day ratio in the surface heat flow is about 1:2 \citep[e.g.,][]{Jaupart:2015}), Earth's mantle undergoes solid state convection with flow rates of cm/year. This convection drives plate tectonics (see \cite{lourenco2026}, this topical collection, and e.g., \cite{sleep_evolution_2015}), a tectonic mode that appears to be unique at the geological present in the Solar System. 
Plate tectonics, wherein the cold and stiff upper layer of seven major rigid plates moving horizontally across Earth's surface participates in mantle convection flow through plate subduction (compare Figs. \ref{fig:tectonic-plates} and \ref{fig:plate vs squishy}), is widely agreed to be of significant importance to Earth's biosphere \citep[e.g.,][]{langmuir_habitable_2012,cockell_habitability_2016, spencer2022biogeodynamics,lingam_stars_2024, stern2016plate}, with feedback from life. It can be argued that its descriptive feature, the rigid plates that make up the lithosphere, is not as important in this respect as  
the associated geochemical 
cycling between the interior and surface reservoirs. For this reason, it is not so important to know when modern plate tectonics exactly began, although estimates range from after the crystallization of the magma ocean \citep[e.g.,][]{korenaga2021} to some time in the Proterozoic less than 2.6 Ga ago \citep[e.g.,][]{cawood_continental_2019}, or as late as in the early Phanerozoic \citep[e.g.,][]{stern_2024}, 0.54 Ga ago (see Fig. \ref{fig:Geological time table} for a compilation of the geological time table as used in this article). What is more important is that earlier tectonic styles, such as squishy--lid tectonics  \citep[e.g.,][and compare Fig. \ref{fig:plate vs squishy}]{lourenco2020}, likely involved similar material cycles and rates of heat transfer. Taken together, these tectonic styles might be better termed ``mobile--lid tectonics'', although ``plate tectonics'' as it is often used in planetary science, means just that. The tectonic cycle produces the Earth's oceanic and continental crusts, 
transfers basic nutrients to the surface, and buffers the mass and composition of the oceans. It has been argued that Mars \citep[e.g.,][]{gregg_tectonic_2015} and Venus \citep[e.g.,][]{harris_crustal_2014, rolf_dynamics_2022}
exhibited some form of mobile--lid tectonics in their early evolutionary histories but currently show no evidence of large--scale horizontal lithospheric movement. Instead, Mars, Mercury and, possibly, Venus are usually cited as examples of stagnant--lid tectonics to describe a tectonic style without large-scale horizontal movement of lithospheric units and very limited cycling. It should be noted that an immobile lid covering a convective layer is the natural outcome of convection in a fluid with strongly temperature--dependent viscosity \citep[e.g.,][]{Schubert:2001}.

Rock, being a multi-component chemical system, partially melts at temperatures above the solidus temperature, releasing buoyant melt that can rise to the surface where it crystallizes as basaltic rock to form the planetary crust. On Earth, because of the near-surface layers 
involved in convection, (hydrated) basaltic crust is subducted into the mantle, where it partially melts in the presence of water and sediments to form granitic crust that builds up into a thick layer of mostly sub--aerial continental crust. The latter is generally less dense than the basaltic crust, is comparatively stable, and rises about 2 km on average
above the sea floor. Where the continental crust has been tectonically deformed (e.g., the Alps and the Himalayas), it can rise up to 10 km above the sea floor. The continents thus cause the Earth's hypsometric curve to have two peaks and mostly rise above the ocean waters that naturally fill the basins provided by the basaltic crust, hence the term ``oceanic crust". In the Archean, formation of the oldest parts of the continents, the cratons, may have involved melting induced by mantle upwelling and the injection of mafic crust by volcanism. Zircon data suggest the formation of granites 
as early as 4.1 Ga ago \citep[e.g.,][]{Wilde2001,Mojzsis2001,Valley2002}. \citet{chowdhury_continental_2025} argue that the hypsometry evolved through the Archean during which the cratons became sub--aerial. However, it is important to acknowledge the paucity of rock records from the Archean eon, a consequence of continental erosion and crustal recycling. 

The continental crust covers about 40\% of the present Earth's surface and may have been so for most of the geological past \citep[e.g.,][]{chowdhury_continental_2025}. 
Since much of the low-lying continental margins are submerged, the part of the Earth's surface above sea level is 29\% of the total. The division of Earth's surface into oceans and land has been important for terrestrial life since at least the Cambrian \citep[e.g.,][]{awramik2007evolution,simpson_bayesian_2017,lingam_life_2021}. Certainly the water in the oceans is important because life needs water, but most life, at least life as we know it, is directly or indirectly phototrophic -- it lives off the sun's radiation. The importance of land surfaces is illustrated by the large proportion of total biomass that is land-based life. Land-based biomass exceeds oceanic biomass by two orders of magnitude, even though the dead fraction of terrestrial biomass is 2/3 of the total \citep{bar-on_biomass_2018}. Furthermore, terrestrial biomass mostly consists of autotroph producers, whereas much of oceanic biomass (oceanic referring to the deep ocean) contains  heterotroph consumers of terrestrial  biomass. This extends the reach of phototrophy into the dark realms of the deep ocean, where some work shows that there may even be an obliquity dependence when we consider an exoplanet context \citep[e.g.][]{Lerner2025}.

But it is not just the availability of sunlight that favors continental areas. Important basic nutrients such as phosphorus and nitrogen are provided by the weathering of continental rock (e.g., \citealt{glaser_detectability_2020, guimond2026waterland}) and are transported to the oceans by runoff and rivers. Seafloor weathering can also provide nutrients, but at much lower rates, as argued by \cite{lingam_dependence_2019} and \cite{glaser_detectability_2020}, due to higher flow velocities in rivers, increased pH in ocean waters, and the lack of erosion providing fresh, weatherable surfaces.

Having roughly equal proportions of land and oceans is an advantage for a planetary biosphere. As \cite{lingam_dependence_2019} show, this maximizes the availability of both nutrients and water. However, it may not be a natural outcome of the long-term evolution of a plate tectonics planet. 
\cite{simpson_bayesian_2017} argued, based on Bayesian statistics, that the more likely outcome is an Earth entirely covered by oceans but cautioned that  feedback mechanisms may cause a mostly land covered planet. \cite{honing_continental_2016, honing_land_2023} argued that, indeed,  feedback in continental growth and water cycling between the interior and the oceans would make a planet almost entirely covered by continental crust a significantly more likely outcome than an ocean covered planet or a planet with an Earth-like balanced ocean/land distribution. The latter outcome is the least likely, but it would maximize the length of continental arc subduction zones.  
Both land and ocean planets would be habitable, but the net primary productivity\footnote{Net primary productivity (NPP) is the rate at which energy (carbon) is stored as biomass by plants and other primary producers in an ecosystem.} would be only about one percent of Earth's, and the O$_2$ productivity might even be insufficient for oxygen-based life. 

The continental crust cycle of production and erosion is linked to the water cycle; both are part of plate tectonics geochemical cycles. Water is transported in the oceanic crust and its sedimentary cover to subduction zones where much of it is carried to depth within the subducting slab, stored in hydrous minerals and directly in pores. Some of this water is released at depths between 100 and 200 km, where important hydrous minerals such as serpentine break down and release water, contributing to the formation of new granitic continental crust. The water that is subducted deeper is absorbed by the mantle (which is far from saturated with water), reducing the  viscosity of the mantle and thus working against the effect of the decreasing temperature to increase viscosity and decrease the speed of mantle convection. The continental crust is eroded by surface weathering and subcrustal erosion, with much of the sediments subducted with the slab. At present and probably for the Phanerozoic, both the growth and erosion rates of the continental crust and the rates of water de- and regassing have been roughly in equilibrium. It should be noted that the present-day distribution of continental crust surface ages  \citep{goodwin_precambrian_1996} can be qualitatively explained by assuming that crust production is proportional to plate speed and that erosion is proportional to the continental surface area even if the continents formed early and have persisted since then \citep{korenaga2021, guimond2026waterland}. Thermodynamically, an equilibrium would be favored where the length of subduction zones is minimized, as for a planet that is either almost entirely covered by oceans or by land. Instead, the Earth operates with the total length of the subduction zones almost maximized, close to an equilibrium which, as \cite{honing_bifurcation_2019} and \cite{honing_land_2023} have shown, is stable with respect to the water cycle but unstable with respect to the continental crust cycle. In the language of non-linear system dynamics, the Earth operates and evolves close to and along an unstable fix point of the system, maximizing the rate of entropy production.    

It is widely accepted that mobile--lid tectonics effectively cools the interior by cycling cold lithosphere with the deep interior. Thereby, mobile lid tectonics effectively powers the core dynamo. The buoyancy released as the core cools and freezes is directly related to the rate at which the mantle cools the core as we will discuss further below in section \ref{p:magnetism}. 

\subsubsection{Atmosphere and oceans 
}  \label{p:atmosphere and oceans}

We give a brief overview of the history of Earth's atmosphere (see \citealt{Steinmeyer2026} in this collection for a more detailed discussion) and ocean to the extent possible given that there is less and less information available as one moves farther and farther back in time. 

The formation of the Earth's early atmosphere and ocean, which sets the stage for its early habitability, remain a topic of debate and intense discussion (see \citealt{salvador2023magma} for a recent review). The canonical view in recent decades has generally been that after the dispersal of an early nebular atmosphere \citep[e.g.][]{mizuno1978instability,Hayashi1979} a post-accretion magma ocean atmosphere would form. This atmosphere would have consisted mostly of CO$_2$ and/or N$_2$ with large amounts of water vapor -- typically termed a magma ocean steam atmosphere. Within $\sim$10$^{6}$ years the surface and atmosphere would cool sufficiently for the steam atmosphere to condense out on the surface forming Earth's first oceans \cite[e.g.][]{Hamano2013,salvador2017}. In the past few years the composition of the magma ocean atmosphere has come under scrutiny. Models show that these atmospheres can differ greatly depending on  the redox state, C/H ratio, and water inventory of the mantle \citep[e.g.][]{bower_retention_2022,maurice2024volatile,nicholls2024magma}. These magma ocean atmospheres influence both when and how Earth's first oceans formed. They would also influence the composition of the post-magma ocean atmosphere, which could be dominated by either N$_2$, CO$_2$, CO, CH$_4$, or H$_2$O with minor contributions from any of those alongside H$_2$ \citep{bower_retention_2022,maurice2024volatile}. Here it is clear to see how constraints on Earth's magma ocean atmosphere may play into exoplanet atmosphere observations in similar epochs, and perhaps vice versa.

If indeed the canonical picture is correct then a few interesting observations can be drawn from the data and associated models. Firstly, there is support for what has been termed a ``Cool Early Earth" \citep{Valley2002} with liquid water present on Earth's Hadean surface supported by zircon measurements \citep[e.g.][]{Mojzsis2001,Wilde2001,Valley2002}. Although exactly how much of the surface was covered with liquid water (as opposed to ice if the planet were relatively cold) cannot be well established from the available zircon data mentioned above \citep[e.g.][]{korenaga2021there}. At this point we should also point out the effect of ocean salinity on climate and habitability \citep{olson_effect_2022, langmuir_habitable_2012}. Salt in ocean water leads to climate warming (via a lowering of the freezing point of water) but too much salt has negative effects on most terrestrial biology. Yet, the estimates of salinity \citep{marty2018salinity} and pH \citep{krissansen2018constraining} in the Archean are quite poor \citep[e.g.][]{CatlingZahnle2020}.
Regardless, it is not difficult to see that, in the early years of the Hadean era, the Earth's ocean was likely covering a fairly flat surface with limited or no exposed land. Different publications give varying results for the amount of exposed land through time \citep[e.g.,][]{korenaga2017global, cawood_secular_2022,rey2024archean,chowdhury_continental_2025}, and this topic is covered in detail in another article in this collection \citep{guimond2026waterland}. In the short term (the next decades) this is an area where exoplanet observations could provide direct insights that we have yet to glean from Earth's record, whereas for the most part it is Earth that provides insights into exoplanets. 
Hence, if we can make similar observations of Earth-like worlds in their early post-accretion phases (see section \ref{sec:observability}) we may begin to get information on the early ocean/land ratios of worlds in their equivalent early to late Hadean phases.

The exact composition of Earth's Hadean 
atmosphere remains largely unknown. However, there are better constraints for the Archean 
\citep[e.g.][for a recent review]{CatlingZahnle2020}. Assuming the mid-to-late Hadean atmosphere resembles the Archean atmosphere, one would expect an N$_2$ dominated atmosphere with partial pressures of CO$_2$ of up to 10s of percent in a $\sim$ 1 bar atmosphere. These large CO$_2$ partial pressures would be necessary to counteract the lower luminosity of the Sun at that time: $\lesssim$ 75\% of modern \citep[e.g.][]{Gough:1981,claire2012evolution}. The need for higher concentrations of greenhouse gases in the Hadean and Archean eras compared to the present day was originally referred to as the Faint Young Sun Paradox \citep{ringwood1961changes,sagan1972earth}. A variety of other greenhouse gases were proposed, such as CH$_4$ and NH$_3$ \citep{sagan1972earth}. However, \cite{kuhn1979ammonia} showed that NH$_3$ was unstable in the atmosphere of early Earth. Neither can one simply rely on CH$_4$ because, once ratios of CH$_4$/CO$_2$ reach values of $\sim$0.1 or higher, hydrocarbon hazes  form and cool the atmosphere \citep[][]{haqq:2008}.
Other authors have proposed that a small increase in CO$_2$, along with fewer clouds and a resultant lower albedo, may have been sufficient \citep[e.g.][]{Rosing2010,Goldblatt2021}, as opposed to higher CO$_2$ partial pressures, which may not be supported by the geological record \citep[e.g.][]{Rosing2010}. More recent work demonstrates that CO$_2$ amounts in the Archean could have been as much as 2500 times higher than modern levels,  while CH$_4$ could have been  as much as 10$^4$ times higher \citep[e.g.][]{CatlingZahnle2020}. We also have proxies that give us estimates of atmospheric density. A number of geochemical proxies for 3.0--3.5 Ga \citep[e.g.][]{marty2013nitrogen} give atmospheric pressures ranging from 0.5 -- 1.1 bar with CO$_2$ values less than 0.7 bar. Work by \cite{som2012air} analyzing fossil raindrops from 2.7 Ga gave values of 0.51 -- 1.1 bar. The fossil raindrop data in combination with later work \citep{som2016earth} utilizing the size of Archean lava vesicles gave a value of 0.23 $\pm$0.23 bar (2$\sigma$) atmospheric pressure. If  Earth's atmospheric pressure indeed  changed so dramatically through the Archean, it is important for exoplanet researchers to consider its impact on the greenhouse gases required to maintain a temperate climate \citep[e.g.][]{charnay2017warm}.

The Archean atmosphere was likely to be reducing, with significant amounts of methane, although limited by the possibility of haze formation as mentioned. Methanogensis dates back to more than 3.5 Ga \citep{wolfe2018horizontal}, giving a methane source through much of the Archean where O$_2$ level remained below 10$^{-6}$ until $\sim$ 2.4 Ga \citep{CatlingZahnle2020}. With rising levels of  O$_2$ during the great oxidation event \citep{holland2002volcanic,lyons2014rise} (and possibly before, \citealt{kasting2025oscillating}), methane would have been destroyed along with its greenhouse effect and this is often cited as the reason for the ``snowball" Earth near the start of the Proterozoic, 
which has been modeled with 3-D general circulation models  \citep[e.g.][]{Feulner2023} and which could possibly be detected on exoplanets in the future \citep[e.g][]{cowan2011rotational,Herbort2025}.

The atmosphere of the Proterozoic is better understood, but there remain uncertainties especially with respect to O$_2$ abundances with estimates ranging from $<$1\% to $>$10\% present atmospheric level \citep[e.g.][]{lyons2014rise,lyons2021oxygenation,reinhard2016earth,mukherjee2025billion}. There do not appear any recorded large divergencies in surface pressure, with N$_2$ remaining the main component of the atmosphere. O$_2$ is not a radiatively active gas (like N$_2$), and hence is not a greenhouse gas like CO$_2$. However, at larger concentrations it can contribute to line broadening of greenhouse gases  (as $\mathrm{N_2}$ does) and hence increase their effectiveness \citep[e.g.][]{Goldblatt2009}.  
This is taken into account in radiative transfer codes in 3-D general circulation models. Most models show that the land-to-ocean ratio has stabilized and is roughly similar to that of today as shown in e.g., \cite{cawood_secular_2022} and in \cite{guimond2026waterland} in this topical collection). Perhaps the most interesting aspect of the Proterozoic is the rise of more complex eukaryote lifeforms instead of simple single--celled prokaryote organisms like stromatolites or cyanobacteria. While the latter had substantial impacts on the atmosphere (e.g., by the production of methane, a potent greenhouse gas), the former had even more so with the continuing rise of photosynthesis and changes in ocean and land biogeochemistry. These latter changes would have, to first order, influenced surface composition, albedo, and cloud cover that could have direct exoplanet observational consequences \citep[e.g.][]{cowan2009alien}. Additionally, two other large scale events with dramatic climate impacts can be quantified in this epoch: the supercontinent cycle \citep[e.g.][]{jellinek2020ice} and snowball Earth \citep[e.g.,][]{hoffman2009palaeogeographic}. An exoplanet in an Earth-like snowball state will be observationally distinct from that of a more temperate world \citep[e.g.][]{cowan2011rotational}. Snowball states tend to be characterized by extremely high surface albedos, dryer atmospheres and an associated decrease in cloud cover \citep[e.g.][]{hyde2000neoproterozoic}. The supercontinent cycle has also been associated with large igneous provinces (LIPs), as discussed by  e.g., \cite{ernst2013large}. These topics will be discussed in more detail in the paragraphs on the Phanerozoic below.

Finally, we reach Earth's most recent geological epoch, the Phanerozoic ($\sim$ 540Ma to present), which includes the Ordovician and the more well-known Cambrian explosion which eventually gave rise to dinosaurs, reptiles, mammals and, of course, humans with their dramatic influence on Earth's present climate. When non-specialists think about Earth, this is likely the only epoch they are aware of, and hence it is often considered the canonical Earth. Yet, as we've seen in this section, it is only a small slice of Earth's entire history. Oxygen level estimates for the first half of the Phanerozoic tend to be higher than that of the 
Proterozoic, but in the latter half they are consistent with the present level. The Phanerozoic is the period in Earth's history about which we know the most. For example, we know that LIPs have been responsible for most mass extinction events in the Phanerozoic \citep[e.g.][]{wignall2001large} with dramatic environmental consequences \citep[e.g.][]{black2024cryptic} that include the release of gases toxic to life, large increases in greenhouse gases, and associated warming which could melt most surface ice affecting a planet's surface albedo. 

In general, it is widely believed that impactors have played a major role in mass extinction events throughout history. 
However, let us look at the most famous mass extinction example at the End-Cretaceous (the Cretaceous–Paleogene (K–Pg) extinction event, also  known as the Cretaceous-Tertiary (K–T)) $\sim$ 66 Ma. This is probably because of the outsize role of the Chicxulub impactor in the demise of the dinosaurs due to its environmental consequences \citep[e.g.][]{morgan2022chicxulub}.\footnote{Note that it may not have even been the largest recorded impactor in Earth's history \citep[e.g.][]{sleep2014physics}, omitting the Moon forming impact of course \citep[e.g.][]{benz1986origin}.}
However, at the time of the Chicxulub impactor the Deccan Traps were being formed via an extremely large LIP. The dating of the impactor and the LIP are close in time \citep[e.g.][]{sprain2019eruptive,schoene2019u} and hence the on-going LIP may have already been weakening many species before the impactor delivered the \textit{coup de gr\^ace}. Yet, there remain some doubts as to whether an impactor played the major role in the End-Cretaceous mass extinction event \citep[e.g.][]{bond2014volcanism}. It has long been assumed that Jupiter may play a ``protective" role for Earth by preventing asteroids and comets from entering its orbit \citep[e.g.][]{wetherill1994possible}, although recent work has put this into question \citep[e.g.][]{horner2008jupiter,horner2009jupiter,horner2010jupiter}.

It has also been estimated that temporally overlapping large LIPs could possibly drive an Earth-like world toward that of a runaway greenhouse as on modern Venus \citep{way2022large}. Hence, while most astronomers are focused on finding temperate Earth-like worlds in the habitable zones of nearby stellar systems, it could be equally likely that we find more Venus-like worlds for reasons scarcely explored. Earth's climate over the past billion years has oscillated between quite warm climates like that of the Cretaceous and during LIP outgassing, to snowball states as documented in the Proterozoic above. In fact LIPs may even end snowball states faster than otherwise \citep[e.g.][]{lan2022massive}.

Another consequence of a focus on the Phanerozoic is the issue of the carbonate-silicate cycle and the role of modern subductive plate tectonics and volcanism originally postulated by \cite{Walker:1981}. \citet{Steinmeyer2026} in this topical collection discuss this in detail. Here we point out the importance of the carbonate--silicate cycle for weathering, the rate of which depends on temperature and CO$_2$ partial pressure. Nutrient cycles of nitrogen \citep[e.g.][]{Forster2019,Berner2006,Stueken2016}, phosphorus  \citep[e.g.][and references therein]{Alcott2022}, and other cycles \citep[e.g.][]{alcott2019stepwise}, depend on weathering through carbonic acid. 
Walker's original paper speculated that this cycle could have been responsible for stabilizing Earth's climate over the majority of its history, and played a role in addressing the Faint Young Sun Paradox mentioned above. 
While it is uncontroversial to say that Earth has had a form of volatile cycling throughout much of its history, the idea that modern volatile cycling is the same as that of the late Hadean or early Archean is probably misplaced. First, the amount of subaerial land at that time is the subject of intense debate, leaning towards there being 
less than today. This would  impact the land--based carbonate-silicate cycle. One could only rely on seafloor weathering \citep[e.g.][]{chambers2020effect} in the late Hadean to early Archean. Recent work argues that seafloor weathering is not as effective as land weathering \citep{lee2019framework,glaser_detectability_2020, ouyang2025phanerozoic}. 
In addition, it is questionable 
that Earth's tectonic state has always remained the same as today \citep[e.g.][Figure 17]{cawood_secular_2022}. For example, the Hadean could have been in a 
``heat pipe" state \citep{moore2013}.  
It has even been speculated by \cite{moore2017heat} that ``Terrestrial exoplanets appreciably larger than Earth may remain
in heat-pipe mode for much of the lifespan of a Sun-like star", but this is controversial \citep[e.g.,][]{balmer_diversity_2021}.   
It has been proposed that a plutonic squishy  lid tectonic mode \citep{lourenco2018efficient,lourenco2020} was the dominant tectonic state from the late Hadean to the early Archean and that it may resemble the conditions on modern Venus \citep{rolf_dynamics_2022}. The volatile cycle that occurrs in such a tectonic state has scarcely been investigated although e.g., \cite{foley_carbon_2018} argue that a carbonate-silicate cycle should be possible even with stagnant lid tectonics.
In the end, we should be cautious in assuming that all Earth-sized terrestrial worlds have volatile cycling like that of modern Earth. It could be that many or even most are in a Venus-like plutonic squishy lid, a Mars-like stagnant lid, or even a Io-like heat-pipe mode with consequences on their ability to affect any long-term volatile cycling, even if shorter-term means have been proposed \citep[e.g.][]{honing2021early}.

\begin{figure}
    \centering
    \includegraphics[width=0.9\linewidth]{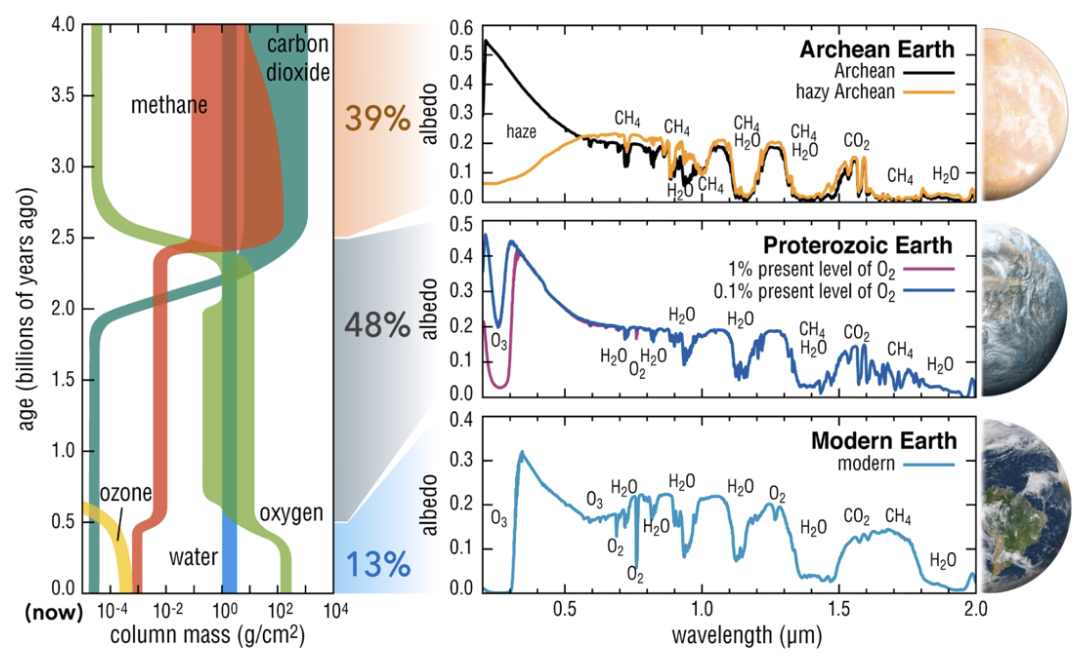}
    \caption{Earth's planetary spectra through time as seen with the LUVOIR concept mission \citep{luvoir2019luvoir}. Credit: G. Arney, S. Domagal-Goldman, T. B. Griswold (NASA GSFC)}
    \label{fig:earth-through-time}
\end{figure}

These three main epochs in Earth's history can most clearly be broadly demonstrated in simulated transmission spectra for the LUVOIR concept study \citep{luvoir2019luvoir} as shown in Figure \ref{fig:earth-through-time}. For example, comparing the Archean and Proterozoic, the Archean does not show the most commonly assumed biosignature gases O$_2$ and O$_3$, since the Archean had so little O$_3$ which is produced from O$_2$. Comparing the Archean to the Proterozoic and modern Earth we see that CH$_4$ is unsurprisingly more prevalent. When we compare the Proterozoic to the present day, we see that the higher abundance of O$_2$ in the atmosphere today results in a much larger O$_3$ absorption feature. This is one of the reasons why the LUVOIR mission concept was designed to be sensitive to such short wavelengths and is expected to be adopted for the Habitable Worlds Observatory \citep{clery2023future}.

\vspace{1cm}

\subsubsection{Magnetism } \label{p:magnetism}
The importance of the Earth's magnetic field in sustaining the biosphere is a debated topic. Certainly, evolved life has learned to use the field for navigation and other purposes. However, it is unclear whether  a magnetic field is an essential element of (eu)habitability. Some argue  that a sufficiently strong magnetic field can protect the atmosphere from  solar (stellar) wind erosion and shield the biosphere from harmful cosmic rays (as discussed in \cite{kubyshkina2026} in this topical collection), making it an essential requirement for habitability \citep[e.g.,][]{lammer_what_2009, mcintyre2019planetary}. However, critical debates about this paradigm have emerged in recent years. These debates are based on spacecraft observations showing similar atmospheric loss rates on Solar System planets despite their differing magnetic field properties and on model calculations \citep[e.g.,][]{blackman_mass_2018, gunell2018intrinsic, ramstad_do_2021, way_synergies_2023}. The role of a magnetic field in protecting a planet's surface from harmful radiation has also been questioned  \citep[e.g.,][]{lingam2019physical}, while the importance of a thick atmosphere has been emphasized \citep{griessmeier_galactic_2016}.

Scaling laws have been proposed for predicting  magnetic field strength based on observable planetary parameters. A critical review of the scaling laws and a comparison with numerical dynamo calculations is  provided in \cite{Christensen2010}. Accordingly, the most convincing scaling law for the magnetic field strength 
was found to be independent of the planetary rotation rate but dependent on the energy flux in the dynamo region available to balance the ohmic dissipation:

\begin{equation}
B \sim
 R_c^{2/3}q_c^{1/3},
\label{eq:field strength}
\end{equation}

\noindent where $B$ is the magnetic induction, $R_c$ is the core radius (for a terrestrial planet), and $q_c$ is the  buoyancy flux in the core. For a thermally driven dynamo, $q_c$ is proportional to the heat flow from the core. For a dynamo driven by chemical convection, $q_c$ is proportional to the rate of growth of the inner core.  Depending on how the dynamo is driven, the efficiency of converting energy into magnetic field energy, and whether the field has a strong dipole component or is largely multipolar, the constant of proportionality in the above equation takes values between 0 and 1 (see \citealt{Christensen2009} and \citealt{Christensen2010}, for discussion). In the latter case, the field strength is only about half that of a strong dipole field. A thermally driven dynamo would have a field strength that is only a fraction of that of a chemically driven dynamo. 

The topology of the field has been found to depend on the strength of the inertial forces relative to the Coriolis force in dynamo experiments and thus on the rotation rate. This is particularly relevant for tidally locked exoplanets, where the rotation rate is equal to the orbital frequency, which is much smaller than Earth’s rotation rate, for example. However, these planets may still have substantial magnetic fields, albeit multipolar,  that could protect them from radiation and stellar winds if their core and interior evolution is similar to Earth's. 

Both terms on the right-hand side of Eq. \ref{eq:field strength} depend on the mass of the planet under consideration. Therefore, it is conceivable that there is a lower mass limit for rocky bodies to sustain a self-generated magnetic field. However, \cite{bryson2015long} use paleomagnetic data from Pallasites to suggest that asteroids with a radius of 400 km could have thermally and chemically driven dynamos. Furthermore, it has been demonstrated that asteroids with radii of tens of kilometers could possess differentiated iron-rich cores and support early dynamos if they accreted rapidly enough relative to the 0.7 Ma half-life of $^{26}$Al \citep{neumann2013thermo}. At the other extreme, while some suggest that super-Earths should cool efficiently and have magnetic fields \citep[see e.g.,][for a discussion]{balmer_diversity_2021}, others have argued that the pressure effect on mantle rock rheology may prevent deep mantle convection, efficient core cooling, and core dynamo action \citep[e.g.,][]{Stamenkovic:2012}. Notably, Mars once had an intrinsic magnetic field, as evidenced by remnant crustal magnetism \cite[e.g.,][]{Acuna1999}. Mercury and Ganymede both have intrinsic magnetic fields \citep{anderson2010magnetic, kivelson1996discovery}. But Venus, which is a similar size to Earth, has no magnetic field. Some believe that any remnant crustal magnetization could be detected  if Venus once had a magnetic field \citep{ORourke2019}.

\subsubsection{Life}\label{subsec:Life}
Despite all efforts to date, 
life beyond Earth has not yet been detected. In the Solar System, primitive carbon-- and water--based life may be detectable at certain depths in Martian soil, where the ambient temperature would permit the presence of liquid water. Similar favorable conditions may exist  in the subsurface oceans of Europa and Enceladus. Exotic life forms that utilize solvents other than water may be discernible on Titan's surface or within the clouds of Venus \citep[e.g.,][]{Schulze-Makuch_life_2018}. These worlds offer urable and habitable niches in which life may have originated and evolved. Considering the Earth, one may inquire how urable and (eu)habitable conditions may have \emph{determined} the origin of life and its evolution to technological intelligence, and to what extent life was subject to \emph{contingency}. In other words, would life on an exoplanet with similar properties invariably evolve in an analogous manner to that of Earth? 

\begin{figure}
    \centering
    \includegraphics[width=0.7\linewidth]{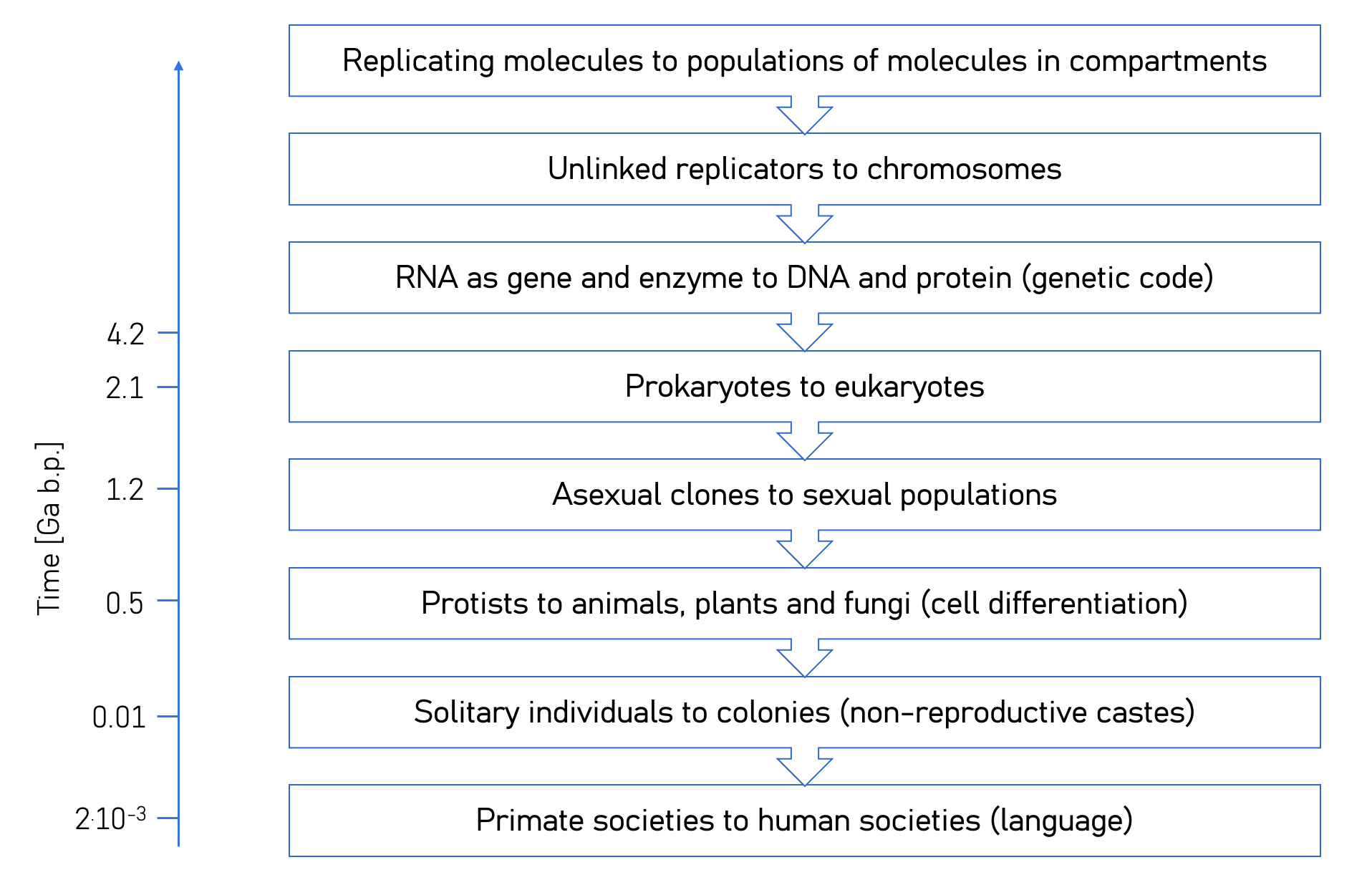}
    \caption{The major transitions in the evolution of life on Earth up to human societies, as compiled by \cite{szathmary1995major}. The first three transitions occurred prior to LUCA. Estimates of when the steps were taken are given}
    \label{fig:evolution}
\end{figure}

Following \cite{smith1995major}, \cite{szathmary2015toward}, \cite{knoll2000directionality}, 
and \cite{henrich2020weird}, 
we briefly outline the major stages of the evolution of life on Earth (compare Fig. \ref{fig:evolution}) and discuss \emph{contingency} and \emph{convergence}\footnote{Contingency refers to an outcome that is decisively influenced by unpredictable circumstances and chance. Convergence, or determinism, describes the tendency for similar forms or solutions to arise independently in different places due to common constraints or similar environmental pressures.} in evolution. It is widely agreed that life originated in the Archean, 
possibly the early Archean, and 
consisted mostly of prokaryotes -- archaea and bacteria. The last universal common ancestor (LUCA), the fundamental node in the tree of life, from which the domains of archaea and bacteria diverged, has recently been dated to 4.2$^{+1.3}_{-1.1}$ Ga b.p. \citep{moody2024nature}. Accordingly, LUCA was an anaerobic acetogen that lived in an established ecological system that it may have supported.  During the Archean era, cyanobacteria or their predecessors invented (oxygenic) photosynthesis and simple forms of multicellularity originated in mats and filaments. The Proterozoic era 
saw the origin of eukaryotes followed by sexual reproduction and the origin of complex multicellularity (animals and plants) during the Neoproterozoic era (1 -- 0.539 billion years ago) or the early Phanerozoic.  
According to the fossil record, life then spread to continental land areas, the Cambrian radiation, followed by the advent of intelligence and eusociality in hominids (living in multigenerational groups with a division of labor) about 2 million years ago. Finally, language, cumulative culture and technological intelligence emerged in humans through culture--gene--environment coevolution \citep[e.g.,][]{durham1991coevolution}.

Following the highly influential work of S. J. Gould \citep{gould1989wonderful}, there has been much debate on whether evolution in general is primarily contingent or convergent. Gould advocated for the primacy of contingency. Others have emphasized the deterministic nature of Darwinian evolution, converging toward the best adaptive solution to an evolutionary challenge \citep[e.g.,][]{morris2003life}. \cite{blount2018contingency} reviewed the debate, including the proper meaning of contingency and cited \cite{desjardins2011historicity}, who clarified that  contingency is intrinsic to path-dependent systems. In such systems, multiple possible paths from an initial state exist that lead to multiple possible outcomes with  ``probabilistic
causal dependencies” linking  the two. \cite{blount2018contingency} further reviewed substantial laboratory research  and paleontological evidence on the subject, finding evolution to be both contingent and convergent. Close lineages tend toward similar adaptations, while distant lineages find distinct solutions. This resembles the dependence of deterministic chaotic systems on initial conditions, as described by e.g., \cite{lorenz1995essence}.

The question of whether evolution is contingent or convergent is particularly relevant to the astrobiology of exoplanets because convergence could imply that advanced life is widespread throughout the universe \citep[e.g.,][]{de2011life, morris2011predicting}. However, \cite{blount2018contingency} have observed that the island of New Zealand lacks native terrestrial mammals, thus arguing against convergence. If contingency were dominant, it would raise doubts about the widespread existence of life, since 
major evolutionary transitions have only occurred once in geological history \citep[e.g.,][]{lingam_life_2021}. Therefore, exploring exoplanets for signs of life is of great importance -- the ``next best thing to rewinding the tape of evolution", to quote \cite{powell2020contingency}. It is interesting to note that \cite{powell2020contingency} acknowledges contingency in evolution, yet argues for the convergence of the evolution toward complex minds if complex bodies were common.  
Assuming convergence in the evolution toward complex minds led  \cite{lingam_life_2021} to suggest that technosignatures in interstellar space may be a more abundant signs of life than biochemical signatures on planets and moons.  

However, recent anthropological research has cast doubt on the idea of determinism in human cultural evolution \citep[e.g.,][]{henrich2020weird,henrich2021origins}. While geographic and environmental factors may suggest determinism \citep[e.g.,][]{diamond1997guns}, other  contingent cultural factors may date back centuries and have had far-reaching consequences. One such transformation is the shift from kin-based societies to more individualistic yet ultrasocial societies. Where this transformation occurred, it was followed by urbanization, the formation of premodern states, and widespread literacy. This was followed by the further 
 specialization of labor, analytical and scientific reasoning, continuous innovation, the Enlightenment and the Industrial Revolution \citep{henrich2020weird, henrich2021origins}.

Even if determinism were inherent to biological and cultural evolution, the conditions of (rocky) planets 
may offer  enough contingency to alter the evolutionary path. 
\cite{chopra2016gaian} and \cite{lineweaver2025other} even argue for a ``Gaian bottleneck" that a biosphere must pass through very early on to avoid extinction due to runaway greenhouse cooling or heating before it can stabilize the climate. Earth may have simply been fortunate enough to pass the bottleneck. Moreover, consider the snowball Earth glaciations during the Proterozoic and the five major mass extinction events in the Phanerozoic.  Various causes of mass extinction events are discussed \citep[e.g.,][]{bond2017causes}, including LIPs and major impact events such as the Chicxulub event \citep{alvarez1980extraterrestrial}. These events caused major climate and ocean chemistry perturbations and consequently mass extinctions, as discussed above in section \ref{p:atmosphere and oceans}.

On a less dramatic scale, some argue that continental drift drove the diversification of lineages and the current distribution of biodiversity \citep{pellissier2018global,gerya2022numerical}. There is consensus regarding the feedback between the evolution of the biosphere and the planetosphere as they co-evolve \citep[e.g.,][]{spencer2022biogeodynamics, stern_2024}. Perhaps the most prominent example of this is photosynthesis and the subsequent oxygenation of the surface and crust, which caused its own mass extinction. Additionally, the increased weathering rate due to the terrestrial biosphere \citep[e.g.,][]{ouyang2025phanerozoic} in the Phanerozoic may have significantly impacted the growth of continents and modern plate tectonics as argued in modeling studies of continental growth by  \cite{honing_continental_2016, honing_land_2023}. In essence, even if biological evolution were entirely deterministic -- which it is now considered false \citep[e.g.,][]{blount2018contingency} -- randomly timed environmental events in the planetosphere will cause contingency. These events are caused by deterministic chaos in nonlinear interior, Solar System, and galactic processes. Would we be writing this article if the Chicxulub event (or the Deccan Trap LIP) had not wiped out the dinosaurs? 
	
\paragraph{Implications for the detection of extraterrestrial life}\label{implications}
    Contingency in evolution, particularly cultural evolution, introduces an additional challenging factor that is nearly impossible to estimate when considering the likelihood of detecting intelligent life beyond Earth (see also \cite{basalla2006civilized} for a critical discussion, in particular, on the perspective of communicating with extraterrestrial lifeforms). It is estimated that there are about $10^{11}$ stars in the galaxy. According to recent work by \cite{quanz2022large}, \cite{lammer_eta-earth_2024} and \cite{scherf_eta-earth_2024}, 
    fewer than a third of these may have rocky planets in their habitable zones. 
    According to \cite{scherf_eta-earth_2024}, approximately 0.5 to 2\% of stars can host a planet with an Earth-like atmosphere dominated by $\mathrm{N_2}$ and $\mathrm{O_2}$ with 1 to 10\% $\mathrm{CO_2}$. These authors estimate the  probability of these planets actually having such an atmosphere to be about $10^{-4}$. Accordingly, there should be approximately $10^5$ planets in the galaxy that could be considered habitable given the aforementioned atmospheric composition constraint. However, given other constraints, this number could be considerably smaller. It is difficult to say how likely plate tectonics will be, for example. The probability of a favorable land/water distribution was estimated by \cite{honing_land_2023} to be a few percent. Thus, the chances for an Earth-like biosphere may be slim, only 1 in $10^8$! This estimate does not consider the likelihood that such a biosphere would evolve technological intelligence. 
    
    Recognizing that Earth may be rare and that the likelihood of finding a second planet like Earth may be very small does not mean that searching for extraterrestrial planets and life is not worthwhile. Rather, it suggests that we should expand our horizons and look for a greater variety of planets, life forms, and biospheres \citep[e.g.,][]{Schulze-Makuch_life_2018}. 
	
\section{Observability of Phenomena at Interstellar Distances }\label{sec:observability}

The previous sections highlight the variety of potential phenomena on terrestrial exoplanets, and how ``2-D'' and ``3-D'' processes may profoundly impact a planet's long-term habitability. However, interstellar distances limit us to one-dimensional characterization of exoplanets for the foreseeable future, as higher-order characterization requires capabilities far beyond current techniques and technologies \citep[e.g.,][]{Turyshev2025}. Using today's techniques, we can study the upper atmosphere annuli of terrestrial exoplanets with transit spectroscopy and their total thermal emission with secondary eclipses and phase curves. In the coming decades, unresolved direct spectroscopy of the entire planetary disk will enable us to investigate the deep atmospheres, and potentially the surfaces, of terrestrial exoplanets. The challenge will be to find signs of complex planetary processes within these datasets, which lack spatial resolution across planetary surfaces. The opportunity lies in the large and varied set of exoplanets with parameters and conditions beyond those in the Solar System. In this section, we explore how the processes discussed in this article and in this topical collection might be detectable on terrestrial exoplanets in the coming decades (see also \citet{lagage2026} in this topical collection).

\paragraph{Atmospheric composition}
Of all the characteristics that could be investigated in detail for terrestrial exoplanets in the not-too-distant future, atmospheric composition is the most promising. Current facilities can probe exoplanet atmospheres using transit spectroscopy. However, this technique is most effective when the planet-to-star radius ratio is large (i.e., large planets around small stars) and is most sensitive to the upper layers, which may not represent the bulk of the planet's atmosphere. Furthermore, when probing the very small signals associated with terrestrial planets and their atmospheres, spectral contamination caused by the inhomogeneous surfaces of stars (i.e., starspots) introduces a significant source of systematic error that is difficult to overcome \citep[e.g.,][]{rackham2023}. 

Secondary eclipse observations and phase curves have successfully measured thermal emission from small exoplanets around low--mass stars \citep[e.g.,][]{kreidberg2019}. By comparing the measured nightside surface temperature to the dayside temperature, these observations can reveal whether a planet has an atmosphere capable of large--scale heat transport. The  rocky planets in the TRAPPIST-1 system that have been studied with the James Webb Space Telescope (JWST), TRAPPIST-1b, -1c, and -1d appear to lack such an atmosphere \citep{greene2023, zieba2023, piaulet2025strict}. The upcoming Ariel mission will perform transmission and eclipse spectroscopy of approximately 1000 warm and hot planets ranging in size from super-Earths to gas giants\footnote{\url{https://www.esa.int/Science_Exploration/Space_Science/Ariel_factsheet}}. Again, the lower--mass super--Earth--sized planets in the Ariel target list preferentially orbit low--mass stars (M dwarfs); however, it is unknown whether these planets can be habitable. Even if some or all of the super--Earth--sized exoplanets are terrestrial, questions remain about whether rocky planets in the liquid water habitable zones of M dwarf stars can retain significant atmospheres over geologic timescales \citep[e.g.,][]{vanlooveren2025}.

Probing the atmospheres of terrestrial exoplanets around Sun-like FGK stars requires shifting to observations that can more effectively distinguish small planetary signals from starlight contamination. One approach currently used for giant planets leverages the spectral difference between stellar and planetary atmospheres; most stars do not exhibit molecular absorption features. Cross-correlating a high-dispersion molecular model with a star and exoplanet spectrum taken at high spectral resolution has achieved detections of molecules despite large amounts of stellar contamination  \citep[e.g.,][]{landman2024}. This technique can reveal a planet's spin rate and the relative abundance of its atmospheric constituents. However, obtaining absolute molecular abundances is challenging. It is uncertain whether this technique can be used to study the atmospheres of terrestrial exoplanets with the James Webb Space Telescope (JWST) or the future Extremely Large Telescope\footnote{\url{https://elt.eso.org}} (ELT) remains uncertain, though it is likely to be more effective for planets around low--mass host stars.

\begin{figure}
     \centering
     \includegraphics[width=1\linewidth]{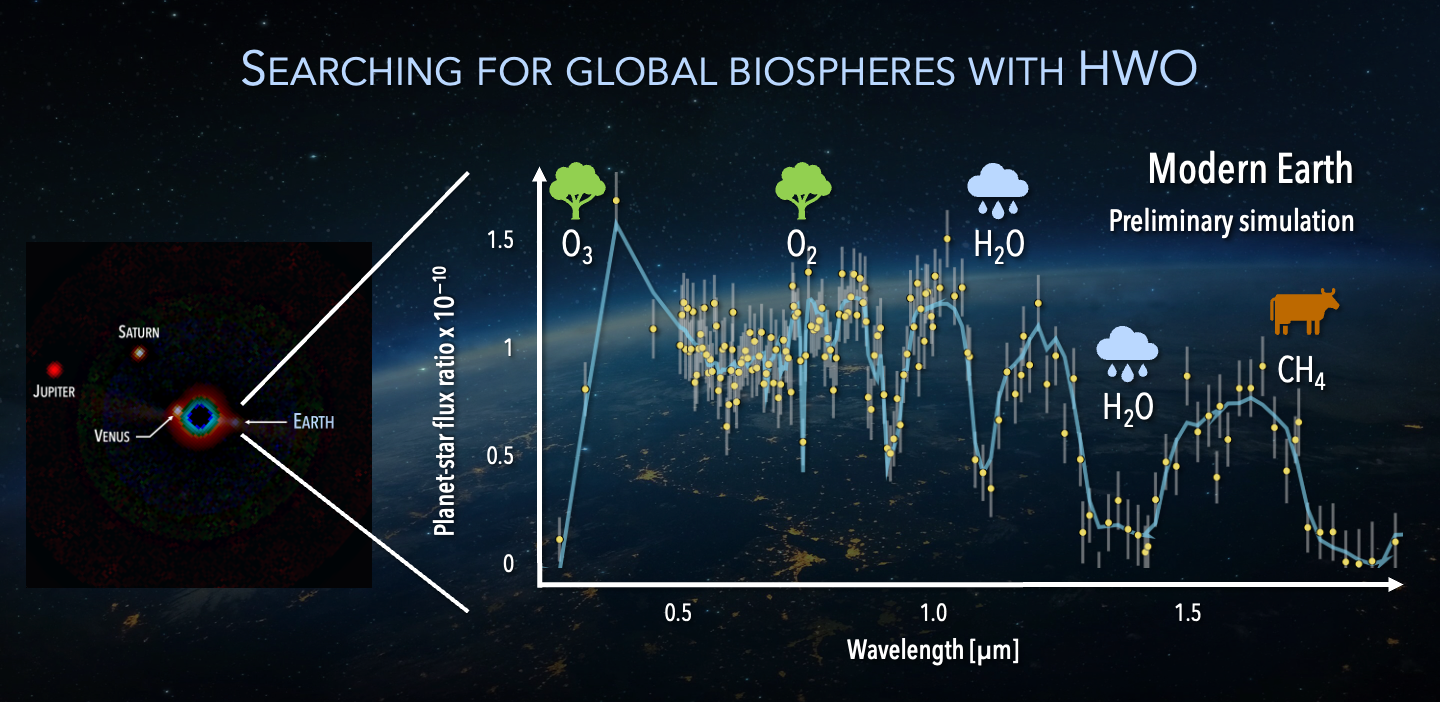}
     \caption{Preliminary simulated HWO high-contrast spectrum of a modern Earth-twin orbiting a nearby Sun-like star. This spectrum is likely of sufficient quality to 
     measure abundances of key indicators of habitability (H$_2$O) and atmospheric biosignatures (O$_3$, O$_2$, and potentially CH$_4$ for earlier phases of the inhabited Earth). The total time to acquire this spectrum, in multiple observations, varies dramatically depending on the distance of the system from the Sun. Credit: J.~Lustig-Yeager (JHU-APL), T.~Robinson (U of Arizona), \& G.~Arney (NASA GSFC)}
     \label{fig:hwo_earth}
 \end{figure}

High-contrast direct spectroscopy, in which the bright host starlight is dramatically suppressed, is likely the best - perhaps the only - way to study the atmospheres of temperate, terrestrial exoplanets around Sun-like stars in the near future. Achieving the extremely high contrast required for such observations (planet-to-star flux ratios of less than  $10^{-10}$ at visible wavelengths) demands a space-based telescope because the Earth's atmosphere will likely limit the achievable contrast, even when using advanced coronagraphs paired with extreme adaptive optics systems on a future Extremely Large Telescope (ELT). The key goal of NASA's Habitable Worlds Observatory (HWO) mission concept\footnote{\url{https://science.nasa.gov/astrophysics/programs/habitable-worlds-observatory/}} is obtaining direct spectra of potentially habitable exoplanets around FGK stars. Figure~\ref{fig:hwo_earth} shows a \textit{preliminary} simulated direct spectrum of the modern Earth around a Sun-twin star as might be obtained with HWO.

\paragraph{Atmospheric and surface biosignatures}\label{biosignatures}
The simulated direct spectrum of the Earth as an exoplanet in Fig.~\ref{fig:hwo_earth} helps us to understand what phenomena we will and will not be able to study in the coming decades. Molecular absorption features of water vapor (H$_2$O), ozone (O$_3$), and molecular oxygen (O$_2$) are readily visible. Methane (CH$_4$) is very difficult to detect at modern Earth concentrations, but could be more detectable at earlier phases in Earth's evolution (the Archean eon; see Fig.~\ref{fig:earth-through-time}). Ozone, molecular oxygen, and methane are all atmospheric biosignatures--gases produced by life at different concentrations over Earth's inhabited history. Thus, signs of a global biosphere are potentially detectable on Earth-like exoplanets around Sun-like stars in the coming decades. 

We emphasize, however, that no molecule is automatically a biosignature in every planetary and stellar context, as demonstrated by the healthy body of literature on potential biosignature false positives. The key to robust atmospheric biosignature identification is to have enough information to reliably model the source and sink rates of various gases in the absence of biology and look for anomalously high molecular abundances that point to an additional source. This will require accurate knowledge of the input stellar spectrum across a wide range of wavelengths, among other planet characteristics. \citet{meadows2022} presents a generic framework to assess life detection claims that, if followed, will increase confidence in exoplanet biosignature detection and interpretation.

The direct spectrum of a temperate terrestrial planet could also contain light reflected from its solid surface. In that case, surface biosignatures such as microbial pigments and/or the sharp increase in reflectivity of photosynthetic vegetation -- the so-called red edge -- could be detectable \citep[e.g.,][]{borges2024}. However, detecting surface biosignatures requires very favorable circumstances: a small cloud--covering fraction and a large portion of the planet's surface covered with life. Furthermore, potential false positives caused by non-living mineral pigments (e.g., iron oxide and iron hydroxide) must be carefully considered and ruled out.

\paragraph{Surface features, magnetic fields and geodetic observables} 
Perhaps surprisingly, it may be easier to detect biosignatures on an Earth-like exoplanet than to confirm the presence of \emph{liquid} surface water. While the large amount of atmospheric water vapor absorption seen in the direct spectrum suggests the presence of liquid surface water, it is not conclusive evidence of oceans. Fortunately, obtaining many such spectra over time may allow one to recover the planet's rotation period and reveal albedo inhomogeneities indicative of surface features such as clouds, continents, and oceans  \citep[e.g.,][]{cowan2009alien, berdyugina_rev_2019}. Data from the Earth-observing satellite DSCOVR \citep{herman2018interactive} have been used to show that even single--pixel observations of Earth (with high cadence) can provide a reasonably good estimate of the land/ocean mask and a planet's rotation rate \citep[e.g.][]{jiang2018using,fan2019earth}. 

Furthermore, the detection of ocean glint, which is caused by specular reflection off liquid water, may be possible. This has been demonstrated using EPOXI satellite observations of Earth \citep{Livengood2011, robinson2011earth}.These observations also suggest that snowball Earth-type scenarios could be observable on terrestrial exoplanets as a wavelength-dependent change in the planet's reflection spectrum compared to modern Earth \citep{cowan2011rotational}. he presence of distinct continents and oceans may provide indirect, tentative evidence of tectonics on a terrestrial exoplanet but is likely only feasible under favorable circumstances, such as a host system that is not too distant from the Sun and a small cloud coverage fraction on the planet.  

Many attempts have been made to infer the presence of magnetic fields on gas giant exoplanets by searching  for auroral emission. Searching for exoplanetary aurorae can take advantage of the fact that certain auroral features of giant planets, such as non-thermal radio emission \citep[e.g.,][]{griesmeier2007} and infrared H$_{3}^{+}$ emission \citep{richey-yowell2025}, should not appear strongly in the spectra of the host stars, thus reducing the likelihood of false positives. Unfortunately, these attempts have not yet been clearly successful. A few tentative detections of radio emission from giant planets have been reported \citep[e.g.,][]{lecavelier2013}, although one of the most promising detections, $\tau$~Bo\"{o}tis~b \citep{turner2021}, was not confirmed in follow-up observations \citep{cordun2025}.

Regarding terrestrial exoplanets, some calculations suggest that radio emission from such planets orbiting M dwarf stars is probably undetectable with current facilities under normal levels of stellar activity \citep{vidotto2019}. However,  others are more optimistic \citep[e.g.,][]{pena-monino2024}. This theoretical debate is unlikely to be resolved without clear observational evidence. Regaerding temperate terrestrial planets around Sun-like stars, it appears that much predictive theoretical work on potentially promising magnetic field signatures remains to be done.

Direct constraints on interior structure of exoplanets beyond mass and radius measurements which result in bulk densities, are scarce \citep[e.g.,][this topical collection]{baumeister2025fundamentals}. 
In the Solar System,  moment of inertia (MoI) measurements are a powerful tool for determining the interior density distributions of planets and moons, allowing estimates of core sizes. The MoI factor can be calculated from measurements of the gravitational moment, $J_2$, from the orbital motion of spacecraft or natural satellites \citep[e.g.,][]{iess2018MeasurementJupiter} or from observations of the precession of the planet's rotation axis \citep[e.g.,][]{folkner1997InteriorStructure}. However, even for Venus the MoI factor is poorly constrained \citep{Margot2021}. Unfortunately, the first type of measurement is not possible for exoplanets, and the second type would  require high-quality direct spectra obtained over extremely long timescales. Closely related to the MoI factor are the fluid Love numbers $k_n$ and $h_n$ \citep[][]{love1911ProblemsGeodynamics}, which describe, respectively, the gravitational response and shape deformation of a planet of degree $n$ under the influence of a perturbing centrifugal or tidal potential. Of particular interest  is the second--degree fluid Love number $k_2$. Assuming that the planet is in hydrostatic equilibrium, $k_2$ depends solely on the density distribution of the interior and $h_2 = 1 + k_2$. Thus, measuring either $k_2$ or $h_2$ can  allow to us to meaningfully characterize the interior structure in principle, as discussed in e.g.,  \cite{padovan2018MatrixpropagatorApproach}, \cite{ baumeister2020MachinelearningInference}, and \cite{ baumeister2023ExoMDNRapid}.

The shape deformation of a planet, which allows for the  inference of $h_2$, can potentially be measured in transit light curves as a deviation from the light curve produced by a perfectly spherical planet \citep{hellard2019RetrievalFluid, akinsanmi2019DetectabilityShape}. If the planet is on an elliptical orbit  and is close enough to its host star, the orbit will precess at a rate determined in part by the planet's gravitational response, thus allowing an estimate of $k_2$ \citep{csizmadia2019EstimateK2}. Currently, $k_2$ has been measured for a few hot Jupiter exoplanets \citep{csizmadia2019EstimateK2, barros2022DetectionTidal, bernabo2024EvidenceApsidal, bernabo2025CharacterisingWASP43bs}. Such measurements will be much more difficult for rocky planets. \citet{kalman2025ProspectsDetecting} estimate that oblateness values $f$ below that of Jupiter ($f\approx0.06$) are not detectable with current transit photometry methods \citep[see also][]{berardo2022EffectsPlanetary}. Therefore, measuring the oblateness of rocky planets, such as Earth with an oblateness of $f\approx0.003$), is likely not possible in the foreseeable future. However, the ever-increasing time baseline of transit and radial velocity measurements may make measurements of Love numbers for sub-Jupiter planets possible in the coming decades.

\paragraph{Technosignatures}
Finally, we consider the detectability of technosignatures, the indirect signs of industry and/or technology. Many phenomena have been suggested as possible technosignatures, including narrowband radio signals \citep[e.g.,][]{tarter1983}; waste heat \citep[e.g.,][]{wright2014}; artificial illumination \citep[e.g.,][]{loeb2012}; and artificial atmospheric constituents, i.e., pollutants such as NO$_2$ \citep[][]{kopparapu2021}. While these phenomena are familiar to us, we must acknowledge that they are extremely recent when compared to the approximately billions of years that life has existed on Earth. Detecting a phenomenon of short duration requires a large sample size. However, without knowing the duration, it is impossible to  determine how large the sample size must be. Furthermore, the proposed technosignatures generally require massive magnifications of the source technology compared to Earth-analog levels to be detectable with current and planned facilities.

The limited, ``1-D'' data that we can collect for exoplanets in the foreseeable future also means that, as with biosignatures, there will be unknowns that complicate the interpretation of possible technosignatures. For instance, if artificial illumination is detectable at all, it could  be mistaken for  the planet's unknown albedo in a secondary eclipse/phase curve observation, or it could be mistaken for a planet with an unusual scattering phase function during a direct observation (a false negative). One must also be wary of astrophysical objects that can mimic technology. A debris disk of interplanetary dust absorbing starlight and re-emitting it at infrared wavelengths, for instance, could be interpreted as a Kardashian Type II civilization giving off waste heat \citep[a false positive;][]{wright2014}. In sum, the prospects for detecting extraterrestrial technology, should it exist in our galactic neighborhood, do not appear promising. However, this possibility should be kept in mind as the characterization of alien worlds grows from a trickle to a flood in the coming decades.

\section{Exoplanet Science Lessons for the Geosciences }
\label{sec:Geolessons}

No subfield of planetary science rivals the wealth of data available for planet Earth. A wide range of geological and geophysical data exists for the planet, including its interior, atmosphere, oceans, biosphere, and space environment. This makes Earth the natural reference point for comparative planetology and space exploration. Space missions exploring the Solar System have revealed the diversity  of its bodies. These include tidally heated moons with enormous heat flow and volcanic activity by Earth standards, ice-covered surfaces with subsurface oceans, diverse atmospheric compositions and surface rock lithologies, varying interior structures, tectonic and volcanic activity, and magnetic properties.   
Space exploration is the foundation of  comparative planetology, a field through which geoscientists can learn to contextualize their home planet. Exploring the galaxy for exoplanets broadens our perspective and reveals even more diversity, particularly because of the large number of planets that will hopefully allow for statistical studies. However, we have come to acknowledge that the Earth is special and that even the configuration of planets in the Solar System does not serve as a universal model.  

The Earth and the planets of the Solar System are near the boundaries or outside the field of the currently known exoplanet populations, as shown in Fig. \ref{fig:exo-populations}. While the radii overlap, it is mostly the orbital distance (and the type of central star) that makes the difference. This is certainly an observational bias that will improve with future observations such as those from the PLATO mission \citep[e.g.,][]{rauer_plato_2014}. 

Of the more than 6000 exoplanets discovered and confirmed  to date (with about 8000 more awaiting confirmation),  the NASA exoplanet archive \url{exoplanetarchive.ipac.caltech.edu/docs/counts_detail.html} (retrieved Dec 22, 2025) currently lists $\sim$1450 planets with radii $\leq$ 2 Earth radii, masses between 0.5 and 10 Earth masses, and estimated densities between 3000 and 6000 kg/m$^3$ that may be rocky. Sixty-seven rocky planets are considered potentially habitable \citep{bohl2025probing}, including the remarkable TRAPPIST-1 planetary system (see \citet{ducrot2026} in this topical collection).

Some Earth-sized planets and most known super-Earths orbit their host planets ``close in'' and are tidally locked in 1:1 resonance. Due to strong insolation and tidal heating, these planets may possess (hemispheric) magma oceans (see \citet{lustig-yaeger2026} in this topical collection). 
These planets provide opportunities to study processes that are thought to have occurred on early Earth \citep[e.g.][]{harrison2020hadean}. Some of these magma--ocean  planets are bright in the infrared and can be observed directly (e.g., 55 Cancri e; \citealt{hammond2017linking}). Assuming the necessary data is available from future observations, other exoplanets
can be used as analogues at other stages of Earth’s history. Similar to the Solar System and supported by planet formation theory, the age of the central star, as determined by stellar seismology, is considered equal to the age of the planets. This allows us to explore different evolutionary paths and improve theoretical models of planetary interiors, geodynamic regimes, atmospheres, oceans, life, and habitability.
 
\subsection{Diversity of composition, mineralogy, and interior processes}

Transit and radial velocity exoplanet detection methods \citep[e.g.,][]{deeg2024transit, trifonov2024radial}, and  Transit Timing Variation observations \citep[e.g.,][]{agol2025transittiming} can be used to infer a planet’s radius and mass, respectively, or both (see Fig. \ref{fig:mass-radius}). From these values, one can estimate the planet’s density and model its internal structure using, in addition, mineral physics data as discussed in \citealt{baumeister2025fundamentals} in this topical collection. These models are aided by 
stellar spectroscopic data that provide constraints on the  composition of the protoplanetary disk and can be used to infer a planet's composition, within the limit of available thermodynamic databases \citep[e.g.,][]{hinkel_stellar_2014, putirka_compositional_2021, unterborn_inward_2018,
wang2022model,
spaargaren2023plausible}. Furthermore, polluted white dwarfs can provide valuable information on the chemical composition of exoplanetary systems \citep{xu_exogeology_2021}. 
Notably,  the current data significantly expand the range of compositions and mineralogies known in the Solar System, in both the quartz-poor 
and the quartz--rich directions \citep[e.g.,][]{putirka_compositional_2021}.

Rocky exoplanets are often assumed to have  characteristics similar to those of Earth and the other planets in the Solar System. Specifically, they are thought to have a structure consisting of a metallic iron-dominated core, a silicate mantle, and a basalt--like crust.  
However, some exoplanets have densities close to that of iron  suggesting high proportions of heavy elements and large cores (see Fig. \ref{fig:mass-radius}). These are  termed super-Mercuries\footnote{For being presumably close in composition to, but larger than, Mercury}. Others have densities between MgSiO$_3$ and water ice, suggesting a high content of volatiles, and are termed super-Ganymedes\footnote{For being presumably close in composition to, but larger than, the Jovian moon Ganymede}. Sub-Neptunes have low densities and are thought to have Earth-like rocky cores and thick atmospheric envelopes. Finally, super-Earths, which were already referred to above,  are planets with Earth-like densities but radii a few times larger. This wide range of planetary sizes and masses results in vastly different properties regarding pressure, temperature, thermal and geodynamic evolution, and, possibly, ultimately habitability  \citep{balmer_diversity_2021, Dorn:2018}. For instance, the pressure in the deep mantle and core of a super-Earth can reach values of terapascals (1 TPa = 1000 GPa) and the temperature can exceed 5000 K \citep[e.g.,][]{Stamenkovic:2012}. At such high pressures, minerals can crystallize into highly compact structures with significantly different physical properties than the lower-pressure phases that dominate the Earth's mantle \citep[e.g.,][]{dutta2022ultrahigh, zurkowski2022synthesis, stamenkovic2011thermal,karato_super_Earth_2011}. These properties may affect mantle convection, thermal evolution, and  core dynamo action. 

\begin{figure}
    \centering
    \includegraphics[width=0.9\linewidth]{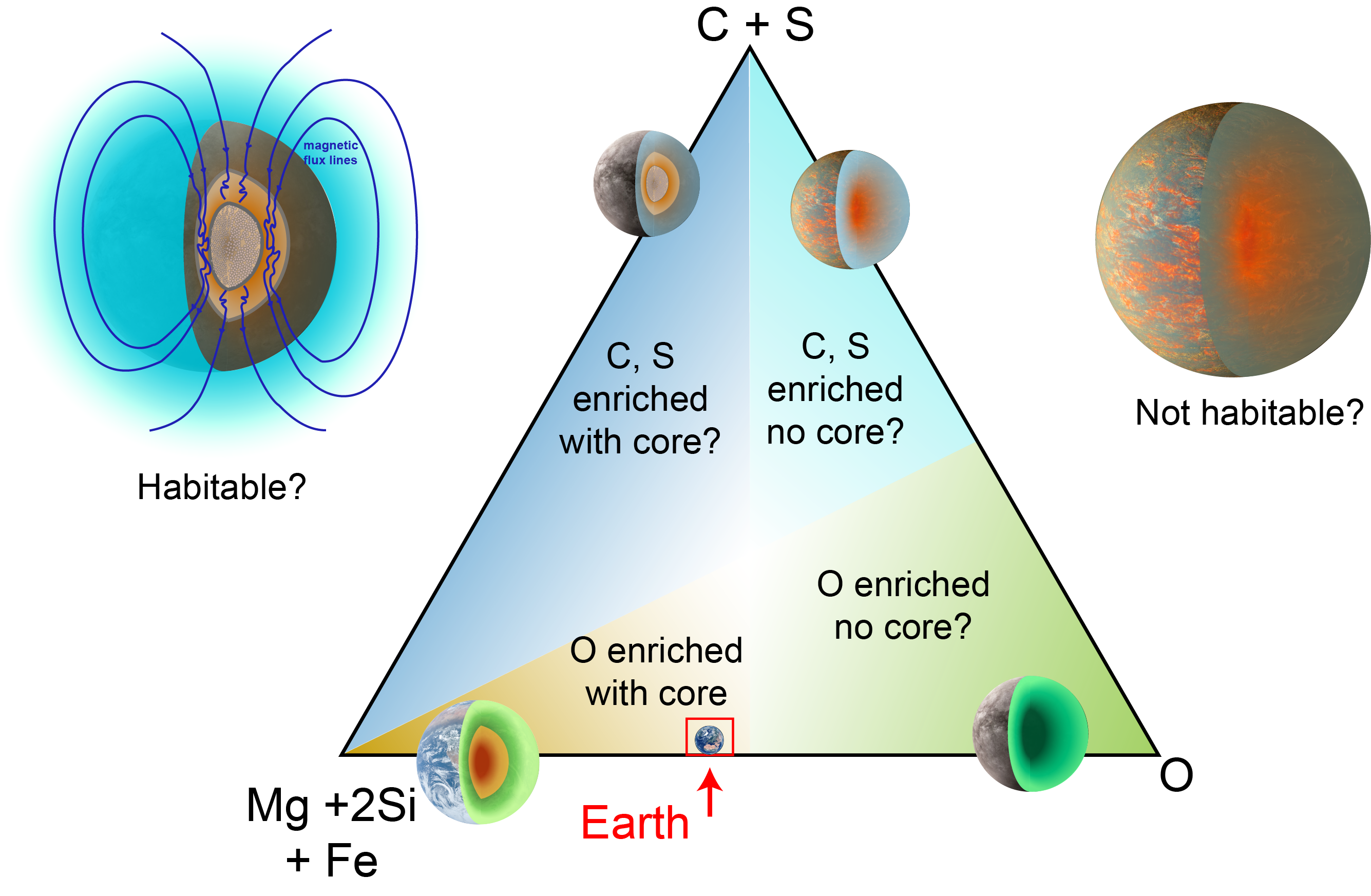}
    \caption{Schematic view of the compositional space spanned by the three major planet forming elements: 1) magnesium, silicate and iron, 2) oxygen,  and 3) carbon and sulfur.(Modified from \citep{unterborn2014role}.) The composition of Earth is also shown. Differences in composition can lead to significantly different interior structures, as oxygen-bearing minerals are replaced by carbon- and sulfur--bearing minerals. Differences in the physio-chemical properties of minerals can lead to significant difference in a planet's large--scale properties. Planets near the Mg+2Si+Fe-rich apex will likely differentiate and form a core, resembling Earth. O-- and C+S--
    rich planets may be undifferentiated, but they have received little study.}
    \label{fig:carbon planet}
\end{figure}

The interior chemistry of a planet also plays a pivotal role in its evolution.  
In recent years, observations of carbon-enriched stars have prompted numerous studies on carbon-enriched planets \citep[e.g.,][]{nisr2017thermal, daviau2018high}, some of which have been expanded to include sulfur as well \citep[e.g.,][]{hakim2019mineralogy}. Interest in studying mineralogies with abundant carbon and sulfur also stems from spacecraft observations of  Mercury and Venus. In oxygen-poor environments, the major elements that form planets can bond with carbon and sulfur rather than with oxygen, resulting in reduced carbon- and sulfur-bearing phases. These phases have also been observed in certain geological contexts on Earth, as well as in extraterrestrial materials such as presolar grains and meteorites. Chemical  diversity can lead to substantial diversity in  planetary properties and interior structure as illustrated in Fig. \ref{fig:carbon planet}. See, for example, \citet{unterborn2014role} and \citet{van_hoolst_exoplanet_2019}.   

 Carbon and carbon--bearing phases such as silicon-carbide (SiC) are less dense than typical mantle silicates. Therefore, in the presence of a mixed mantle -- as suggested by \cite{schmidt2014natural} for reducing conditions -- it would be possible for these minerals to upwell and form a carbon--rich upper layer. A similar mechanism has been proposed as the origin of the floating graphite crust on Mercury \citep[][]{peplowski2016remote, keppler2019graphite}. \cite{miozzi2018equation} conducted numerical simulations of an archetypal SiC mantle and demonstrated that SiC's distinct properties relative to silicates render the mantle less susceptible to convection. However, the simulations do not depict  layered convection, as suggested by \cite{nisr2017thermal}. Unfortunately, the potential dynamic regime of reduced planets has received little study and many of the characteristics of these potential mantles remain unknown. 

Sulfur-bearing materials have received more attention as potential components of Mercury's mantle. For instance, \cite{pommier2023experimental} report the formation of Mg-S and Ca-S bonds in glasses under highly reducing conditions, while\cite{mouser2021experimental} investigate end-member scenarios for mantle evolution based on experimentally collected viscosity data for a Mercury-like magma ocean. Changes in the ratio and chemistry of a planet's major elemental components would affect its composition and, in the extreme scenarios, whether it has a core. The Fe-S-C system exhibits an immiscibility gap at lower pressures, leading to the formation of two distinct metallic phases \citep[e.g.,][]{hakim2019mineralogy}. However, the evolution of the ternary system at pressures and temperatures relevant to Earth-sized planets, 
has not been investigated. Understanding core composition and structure is fundamental to assessing the possibility of a geodyamo (see \citet{lourenco2026} in this topical collection). 
 
Directly imaging the surface of exoplanets with sufficient resolution to distinguish geodynamic regimes will be difficult in the foreseeable future. However, we may be able to infer their tectonics using indirect observations, such as  their atmospheric compositions and potential magnetic fields. Gases released into the atmosphere by volcanism provide insight into a planet’s geodynamics. On Earth, for example, the composition of the atmosphere is a consequence of the (bio)geochemical cycles maintained by the current regime of plate tectonics \citep{schaefer2021air}. Advanced Earth System modeling tools will improve our understanding of the range of possible geodynamic regimes and those that  Earth has passed through. Improved observations of Earth-like planets at different stages of evolution may also offer direct insights into Earth's past and future.
Coupling observations with theoretical models may improve our understanding of planetary geodynamics and the conditions necessary for plate tectonics. Larger planets may be able to retain more heat resulting in more vigorous mantle convection. However, higher thermal energy could potentially prevent the formation of a strong lithosphere. This suggests that a delicate balance is needed to maintain plate tectonics \citep[e.g.,][]{oneill2007, korenaga2010likelihood, foley2012conditions, noack_plate_2014}. Other factors, such as intense solar radiation and the presence orabsence of water (and even life), are thought to play a fundamental role in  permitting or preventing plate tectonics. 
Examining the library of exoplanets will allow us to improve our understanding of how wide (or narrow) this balance is. This will be key to understanding why Earth developed plate tectonics and the main factors controlling it.

 \subsection{Atmospheres and oceans}
The study of exoplanets reveals a wide range of atmospheric compositions and  potential climates \citep{schaefer2021air}. These compositions depend on a planet’s mass, composition, geodynamic regime, and external processes, such as photodissociation and atmospheric erosion. The range of orbital parameters and planetary sizes also strongly affects their atmospheric circulation and climatic patterns. Studying Earth-like exoplanets in various stages of evolution improves our understanding of processes such as the formation of primitive atmospheres, atmospheric escape, and the formation of clouds and hazes \citep{zahnle2017cosmic,Fauchez2019}. James Webb Space Telescope observations will also contribute to our understanding of photochemical processes within exoplanetary atmospheres, which are  key to identifying biosignatures.

Data from transit spectroscopy and secondary eclipses will  reveal a variety of atmospheric compositions and improve our climate models \citep{komacek2021constraining}. Investigating a wider range of conditions will help us assess the robustness and predictability of our current global circulation models of past and future Earth \citep[e.g.,][]{way2021climates}. This information is crucial for understanding how the Earth’s atmosphere and climate have changed over time and what future pathways may be possible. Studying exoplanets that have undergone runaway greenhouse effects or lost their atmospheres will help us identify critical thresholds that trigger these extreme (and potentially irreversible) climate states \citep[e.g.,][]{chaverot2023first,airapetian2020impact,way2020}. Additionally, since exoplanet atmospheres will be our only ``direct" window to their (bio)geodynamics for the foreseeable future, understanding how they operate is crucial to improving our knowledge of planetary geodynamics with sufficient precision to apply it to the past and future of Earth.
 
 Studies of exoplanets have suggested the presence of exo-oceans in the form of subsurface oceans in ice worlds and  deep, global surface oceans in aquaplanets \citep[e.g.,][]{cowan2009alien,cowan_water_2014, piaulet2023evidence}. Observing exo-oceans  will provide new insights into our own ocean, particularly how it formed and evolved (see section \ref{sec:observability}). There are uncertainties about the sources of  water that led to the formation of the Earth’s oceans \citep[e.g.,][]{karato_water_2015}. For example, we do not know how much water primitive Earth had and how much was brought from the outside by comets or chondrites \citep[e.g.][]{marty2012origins}. Future direct observations of magma ocean worlds could provide valuable information about degassing processes and interactions between a magma ocean and the atmosphere \citep[e.g.][]{way_synergies_2023,salvador2023magma}.
Furthermore, observing exo-oceans with different depths and compositions (e.g., ammonia and hydrocarbons) may help us understand the likelihood of forming an Earth-like ocean and maintaining it. Understanding how ocean diversity impacts geochemical cycling, climate, and plate tectonics is also important. This knowledge will improve our understanding of how these processes have operated throughout Earth’s history. For now, part of this work must be based on modeling, but future observations will be crucial to validating these models. 

\subsection{Life}
The search for life beyond Earth remains the primary motivation for extraterrestrial research. This applies to searches within and beyond the Solar System.   Life on Earth can exist in a wide range of conditions, from  deep within the crust to high in the atmosphere\footnote{Self-replication of life in the atmosphere has yet to be observed \citep[e.g.][]{tastassa2024aeromicrobiology}.} \citep[e.g.,][]{Schulze-Makuch_life_2018} and it comes in a multitude of species. However, extraterrestrial life is expected to offer its own surprises. As we concluded in previous sections of this article, finding  a second Earth with \textit{life-as-we-know-it} appears challenging if not unlikely. The first inhabited planet discovered will likely not be another Earth! While this conclusion complicates the search for biosignatures,  a positive result would be all the more exciting and rewarding for our understanding of life on Earth.   
 
\section{Concluding Thoughts on the Perspectives of Exo-Geoscience}\label{sec3}

Exo-geoscience is an emerging field that brings together geoscientists, astrophysicists, planetary scientists, and astrobiologists \citep[e.g.,][]{unterborn2020exogeoscience, shorttle_geosciences_2021}. Geoscientists play a pivotal role in helping to interpret exoplanet observations by offering insights from Earth. In the process, they gain valuable insights about Earth by comparing it to discoveries on other worlds. 
Earth and the other rocky planets and moons in the Solar System continue to serve as \textit{the} reference case for potential discoveries in other planetary systems throughout the galaxy and beyond. However, it is important to understand that the Solar System does not represent a unique pathway for planetary evolution. Due to the dearth of data that will likely persist into the foreseeable future,  we must be very cautious when drawing conclusions about extrasolar terrestrial planets.

As we have noted, Earth is a living planet that has co-evolved with life, possibly after passing through early bottlenecks for life by chance \citep{lineweaver2025other}. These bottlenecks may have been climatic or related to early meteoritic bombardment or due to volcanic activity on the planet.  Earth's location at the right distance from the Sun and in the right corner of the galaxy made it possible for life to emerge. One could argue that Venus and Mars may have developed life early on, but that their orbital distances may not have been quite as favorable. Thus, answering the open question of whether there is extant or extinct life on Mars, and less likely perhaps on Venus, is important. However, one could argue that this question can only be definitively answered by detecting life on these planets. Continued failure to detect life may just keep the question open. 

There is good reason to argue that \textit{life-as-we-know-it} on Earth depends on the properties of this planet. These properties include plate tectonics, which maintains clement climate conditions, renews basic nutrients, sustains bio-relevant geochemical cycles, maintains a diverse environment with land and oceans, and a magnetic field. It is also being increasingly recognized that life plays a role in the global functioning of Earth as an inhabited, habitable planet. Considering the evolution of life on Earth, it is becoming increasingly clear that it is not simply the living planet moving through the evolutionary landscape on an optimal path. Rather  the evolution of the biosphere to consciousness and the ability to explore its cosmic neighborhood and answer the questions posed in this article may be due to contingency in both the geological and   cultural evolution. 
However, it is essential to remain open to other realizations of inhabited planets.     

For geoscientists, interest in exoplanetary science is  largely motivated by the large number of planets that may eventually allow for statistical analysis. As we have described above, exploring the Solar System has provided some insight into planetary diversity.  Exploring exoplanets will provide a much wider range of planetary states that can help to  improve our understanding of the Earth \citep{meadows_exoplanet_2018, lammer_what_2009}. Detecting planets in different stages of evolution will offer insights into how Earth may have been in the past and how it evolved, by studying processes that no longer occur on our planet. However, since direct observations are limited, we will need to enhance our theoretical models and laboratory studies of planetary material properties. The databases currently available and used to investigate processes on Earth need to be expanded to account for those phases that are not common or existing on Earth but can still belong to similarly defined chemical systems and are observed in extraterrestrial materials. The pressure and temperature range in which the physiochemical properties of mineral are determined need to be expanded. At the same time, a more comprehensive database of transport properties should be built. The latter are extremely challenging to determine experimentally and require the development of new techniques. Another frontier for experimental work is the study of hydrogen and its interaction with melts, minerals, and metals, representative of the magma oceans and planetary mantles and cores respectively. Mineral properties lie at the base of every process occurring on planets and as such, it is fundamental to have the most complete possible characterization in order to model planetary processes. And a new generation of models must integrate the various components of the planetary system \citep{foley_whole_2016}.  
The current momentum in exoplanet research will certainly result in a much improved understanding of the Earth. Lessons learned from studying exoplanets will help us distinguish the universal properties of rocky planets from contingent ones and the results of chance.

\backmatter

\bmhead{Acknowledgments}
Thanks to Diana Gentry for advice regarding replication of life in the atmosphere. 
MJW supported by NASA's Interdisciplinary Consortia for Astrobiology Research (ICAR), Nexus for Exoplanet System Science (NExSS), and support from the GSFC Sellers Exoplanet Environments Collaboration (SEEC), which is funded by the NASA Planetary Science Division’s Internal Scientist Funding Model.
P.B. is funded by the European Union (ERC, DIVERSE, 101087755). Views and opinions expressed are, however, those of the author(s) only and do not necessarily reflect those of the European Union or the European Research Council Executive Agency. Neither the European Union nor the granting authority can be held responsible for them.
F.M. acknowledges support from the Carnegie Institution for Science and the Alfred P. Sloan Foundation under grant G202114194. 
J.C.D. is supported by the Portuguese Fundação para a Ciência e Tecnologia, FCT, I.P./MCTES through national funds (PIDDAC): UID/50019/2025 and LA/P/0068/2020 (https://doi.org/10.54499/LA/P/0068/2020). J.C.D. also acknowledges FCT for a CEEC Inst. 2018 contract, CEECINST/00032/2018/CP1523/CT0002 (https://doi.org/10.54499/CEECINST/00032/2018/CP1523/CT0002) and funding from the project GEMMA (PTDC/CTA‐GEO/2083/2021, https://doi.org/10.54499/PTDC/CTA-GEO/2083/2021).

\bmhead{Statements and Declarations} The authors declare that they have no competing interests. 

\bibliography{sn-bibliography}

\end{document}